\documentclass[twocolumn,trackchanges]{aastex701}
\usepackage{float}
\usepackage[utf8]{inputenc}
\usepackage{xcolor}
\usepackage{threeparttable}
\usepackage{epstopdf}
\usepackage{verbatim}
\usepackage{subfigure}

\usepackage{graphicx}
\usepackage{amsmath}
\usepackage{ulem}

\begin{document}

\title{\textit{Chandra} and XMM-\textit{Newton} X-ray Study on the Core-collapse Supernova Remnant N63A in an H II Region}

\author[orcid=0000-0001-8892-8759]{Hanxiao Chen}
\affiliation{Department of Astronomy, Nanjing University, Nanjing 210023, People’s Republic of China}
\email{chenhx@smail.nju.edu.cn}  

\author[orcid=0000-0002-4753-2798]{Yang Chen} 
\affiliation{Department of Astronomy, Nanjing University, Nanjing 210023, People’s Republic of China}
\affiliation{Key Laboratory of Modern Astronomy and Astrophysics, Nanjing University, Ministry of Education, People’s Republic of China}
\email[show]{l.sun@nju.edu.cn}

\author[orcid=0000-0001-9671-905X]{Lei Sun}
\affiliation{Department of Astronomy, Nanjing University, Nanjing 210023, People’s Republic of China}
\affiliation{Key Laboratory of Modern Astronomy and Astrophysics, Nanjing University, Ministry of Education, People’s Republic of China}
\email[show]{ygchen@nju.edu.cn}

\author[orcid=0000-0002-3576-441X]{Jianbin Weng}
\affiliation{Department of Astronomy, Nanjing University, Nanjing 210023, People’s Republic of China}
\email{jianbin.weng@smail.nju.edu.cn}

\correspondingauthor{Lei Sun}
\correspondingauthor{Yang Chen}

\begin{abstract}
N63A is one of the brightest supernova remnants (SNRs) in the Large Magellanic Cloud and provides an excellent laboratory for studying SNR evolution in an H II region. 
We present a detailed X-ray analysis of N63A using \textit{Chandra} and XMM-\textit{Newton} observations, combining both imaging and high-resolution spectroscopy. Our global spectral fitting reveals three distinct thermal plasma components with temperatures 
$kT$$\sim$0.3, 0.7, and 1.5\,keV.
Spatially-resolved spectroscopy with \textit{Chandra} shows substantial variations in physical parameters across the remnant, with the eastern lobes of the optical nebula exhibiting significantly higher ionization timescales than the western region. These multi-temperature plasma components indicate that N63A is evolving in a cloudy interstellar medium: 
the low-temperature component arises from evaporated 
dense cloudlets that are engulfed by the blast wave, the intermediate-temperature component represents shocked inter-cloud medium, and the highest-temperature component likely results from 
the shocks reflected by the clouds. We estimate a Sedov age of approximately 4.5 kyr and an explosion energy of about $4\times10^{51}$\,erg. Comparison of the observed abundance ratios and masses of metal elements with the nucleosynthesis models constrains the progenitor mass to approximately 
$20$ $\rm M_{\odot}$, supporting a single-star core-collapse origin for N63A.

\end{abstract}


\section{Introduction}

Supernova remnants (SNRs) are the product of supernova (SN) explosions that mark the final evolutionary stage of massive stars. SN explosions release large amounts of energy and ejecta that interact with the surrounding interstellar medium (ISM) or circumstellar 
material (CSM). 
SNRs provide important insights into the physical and chemical properties of the ejecta and ISM/CSM. This helps us better understand SNR evolution, metal and energy feedback to the galaxy, etc. 
Furthermore, 
SNRs allow us to trace the properties of their progenitor stars and explosion mechanisms. 
SNe often explode in inhomogeneous environments, which will significantly affect the evolution of the SNRs \citep{McKee1977,Cowie1981}.
N63A, one of the brightest SNRs in the Large Magellanic Cloud (LMC) and the first identified to have formed within an H II region \citep{Dickel1993}, is an ideal target for studying SNR evolution in an inhomogeneous medium.

N63A was first identified as a SNR by \citet{Mathewson1964}. Its spatial coincidence with the 
OB association NGC 2030/LH 83 \citep{Lucke1970} suggests an origin from a core-collapse SN of a massive ($>40\,\rm{M_{\odot}}$) star \citep{Oey1996,Karagoz2023}, which is further supported by Fe K$\alpha$ line-center measurements \citep{Yamaguchi2014}. The whole remnant is embedded in the classical H II region N63, where strong stellar feedback from massive stars has produced a highly inhomogeneous and clumpy ambient medium \citep{Dickel1993}. 
The age of N63A is estimated 
as $\sim$2000--5400\,yr, with an X-ray size of $\sim70^{\prime \prime}$  \citep[corresponding to $\sim$18\,pc in diameter at 50\,kpc, e.g.][]{Dickel1993,Warren2003,Karagoz2023}. Notably, \textit{Chandra} observations analyzed by \citet{Karagoz2023} indicate a high explosion energy ($\sim 8.9\pm1.6\times 10^{51}$\,erg), potentially pointing to a hypernova origin.

The remnant exhibits a triple-lobed optical structure \citep{Mathewson1983}. The two eastern lobes show high [S II]/H$\alpha$ ratios \citep{Payne2008}, confirming shock-heated gas, while the western lobe corresponds to a photoionized H II region \citep{Levenson1995}. X-ray analysis 
suggests that the emission of N63A originates from the shocked ISM rather than SN ejecta \citep{Hughes1998,Russell1990}. Recent ALMA observations mapped the molecular cloud distribution and found that the dense molecular clouds have been completely engulfed by the shock waves, but the clouds resist erosion due to their high density and 
short interaction timescale \citep{Sano2019}.

Previous X-ray studies of the SNR mapped physical parameters via spatially-resolved spectral analysis \citep{Warren2003, Sano2019, Karagoz2023}, but fundamental properties such as the remnant's age and the explosion energy remain poorly constrained. While \citet{Reyes-Iturbide2025} recently performed a global analysis using XMM-\textit{Newton} observations, their approach lacked spatial resolution. 
In this work, we present a comprehensive X-ray analysis of N63A taking advantage of the superb spatial resolution of \textit{Chandra}/ACIS as well as the high spectral resolution of XMM-\textit{Newton}/RGS. We also applied differential emission measure analysis aimed at a better understanding 
of the progenitor and the evolution of N63A.

In Section \ref{sec:obs}, we describe the \textit{Chandra} and XMM-\textit{Newton} observations and the data reduction procedures. Section \ref{sec:spec} presents the spectral analysis, including emission line identification, global spectral fitting with multi-temperature and differential emission measure (DEM) models, and spatially-resolved spectroscopy. In Section \ref{sec:discuss}, we discuss the physical implications of our results, including the properties of N63A, origin of the multiple temperature components, and the nature of its progenitor. Our main findings are summarized in Section \ref{conclusion}.

\section{Observations and Data Reduction}\label{sec:obs}
\begin{figure}
    \centering
    \includegraphics[width=0.98\linewidth]{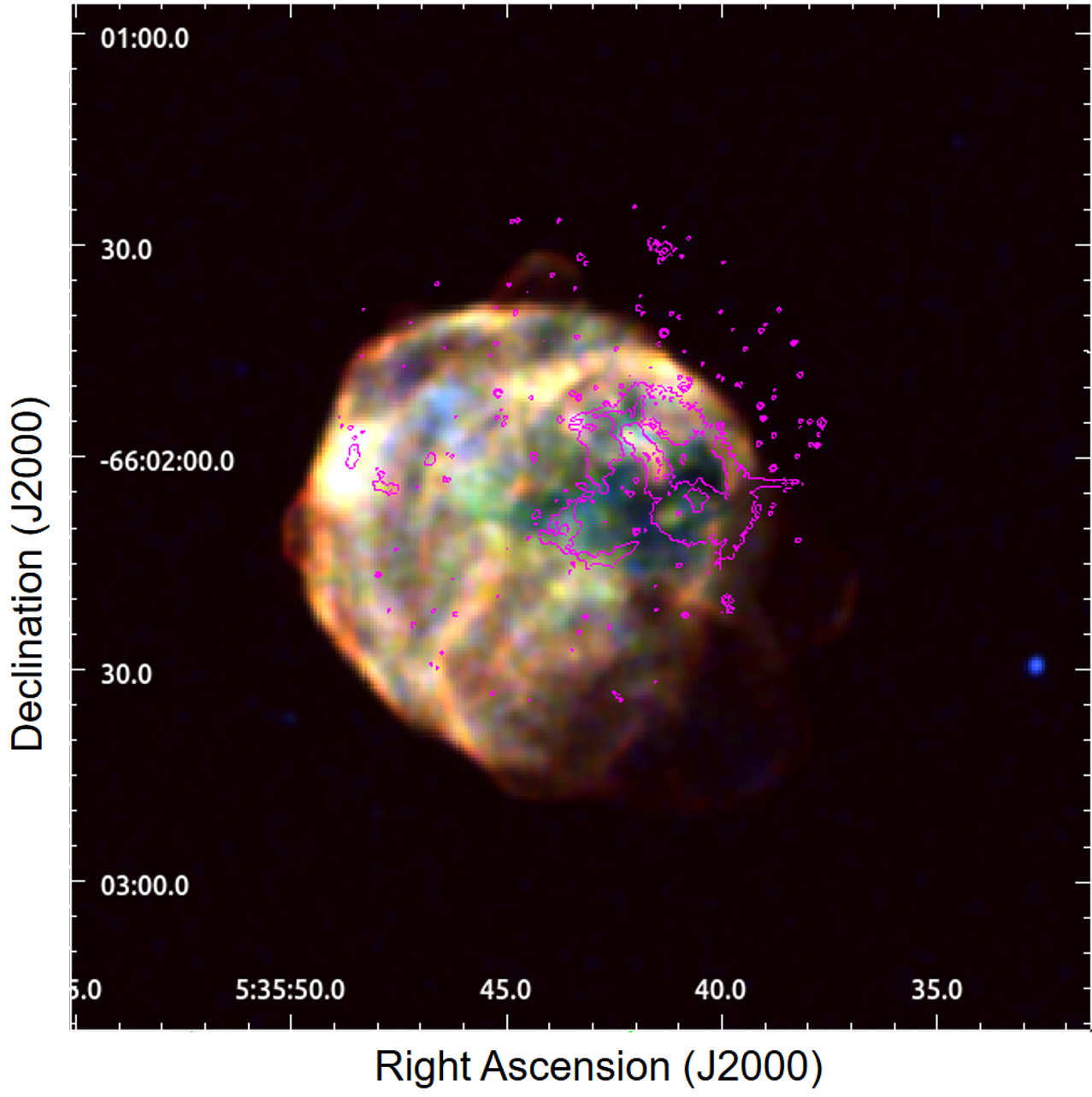}
    \caption{\textit{Chandra} RGB image of N63A  in 0.5--1.2 (red), 1.2--2.0 (green), and 2.0--7.0\,keV (blue) bands. The magenta contours outline the optical nebula.}
    \label{fig:chandra_image}

\end{figure}

\textit{Chandra} observed N63A on 2000 October 16 using the Advanced CCD Imaging Spectrometer S3 array (ACIS-S), under Observation ID 777 (PI: J. P. Hughes).  Data were processed with the Chandra Interactive Analysis of Observations \citep[CIAO, ][]{Fruscione2006} data-processing software (Version 4.16, and CALDB version 4.11.2). All the data were reprocessed with \texttt{chandra\_repro} script. The effective exposure time is approximately 41.0\,ks. We used \texttt{fluximage} to generate broadband and RGB images of N63A (Figure \ref{fig:chandra_image}), and employed \texttt{specextract} to extract source and background spectra. The background region was selected from an annulus source-free region, %
encircling the remnant with an inner radius of 1.5 arcmin and an outer radius of 3 arcmin.

The XMM-\textit{Newton} X-ray satellite observed N63A on 2000 November 15 
(Observation ID: 0109990101, PI: J. Bleekser). Both the Reflection Grating Spectrometer (RGS) and the European Photon Imaging Camera (EPIC; pn and MOS) data were utilized. The RGS, with its high spectral resolution, is sensitive to emission lines in the 0.35–-2.5\,keV range. In contrast, the EPIC-pn and EPIC-MOS have larger effective areas and broader energy range  
up to 10\,keV that provide improved constraints on the high-temperature components. The XMM-\textit{Newton} data were reprocessed using the Standard Source Analysis Software \citep[SAS Version 20.0, ][]{Gabriel2004}. The effective  
exposure times for EPIC-MOS and EPIC-pn are about 9.7\,ks.

\begin{figure}
    \centering
    \includegraphics[width=0.98\linewidth]{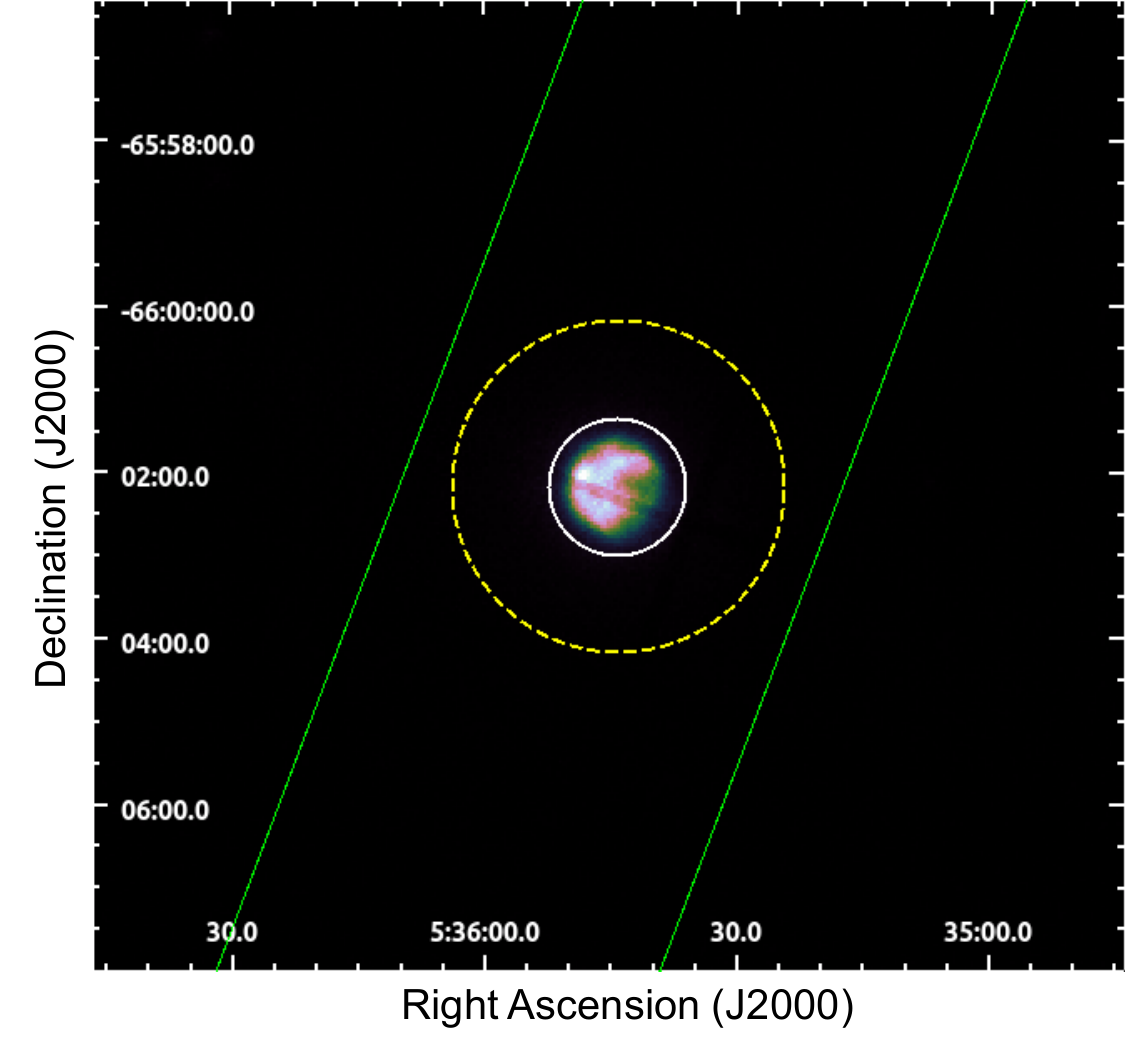}
    \caption{XMM-\textit{Newton} image of N63A. The source region is indicated by a white circle, while the background region is the annular area between the yellow dashed circle and the white circle. The green lines show the RGS dispersion direction and cross-dispersion aperture.}
    \label{fig:xmm_image}

\end{figure}

Following standard procedures, the EPIC-MOS and EPIC-pn event files were regenerated using the SAS tasks \texttt{emchain} and \texttt{epchain}, respectively. Periods of enhanced background due to soft proton flares were identified and excluded by applying the \texttt{mos-filter} and \texttt{pn-filter}, resulting in cleaned event lists for subsequent analysis.

Spectra from selected regions were extracted with the \texttt{mos-spectra} and \texttt{pn-spectra} tasks, which also produce the corresponding response files. The quiescent particle background (QPB) was estimated through the \texttt{mos-back} and \texttt{pn-back} and was subtracted from the source spectra before spectral fitting.

We generated the 
standard RGS response matrices (RMFs) using the 
SAS tool \texttt{rgsrmfgen}, which is optimized for point sources. However, since our target N63A is an extended source, 
its extension in the dispersion direction will lead to confusion
with the wavelength scale. To address this, we applied the FTOOL \texttt{rgsrmfsmooth} to modify the RMF before spectral fitting. The image used for 
the modification is the 0.3--2.5\,keV \textit{Chandra} image of N63A, which has better resolution. After correcting the RMFs, we performed spectral analysis using XSPEC \citep[version 12.14.0, ][]{Arnaud1996} with AtomDB 3.0.9. \citep{Smith2001, Foster2012} 
The EPIC spectra were extracted from the entire SNR region, with background spectra taken from an outer circular region (Figure \ref{fig:xmm_image}).

\section{Spectral Analysis}\label{sec:spec}

In this article, unless otherwise specified, the reference solar abundances used are those 
in \citet{Wilms2000}, and 
the error bars denote 1-$\sigma$ uncertainties.

\subsection{Emission Lines and Their Fluxes}\label{sec:lines}
\begin{figure*}
    \centering
    \includegraphics[width=0.7\linewidth]{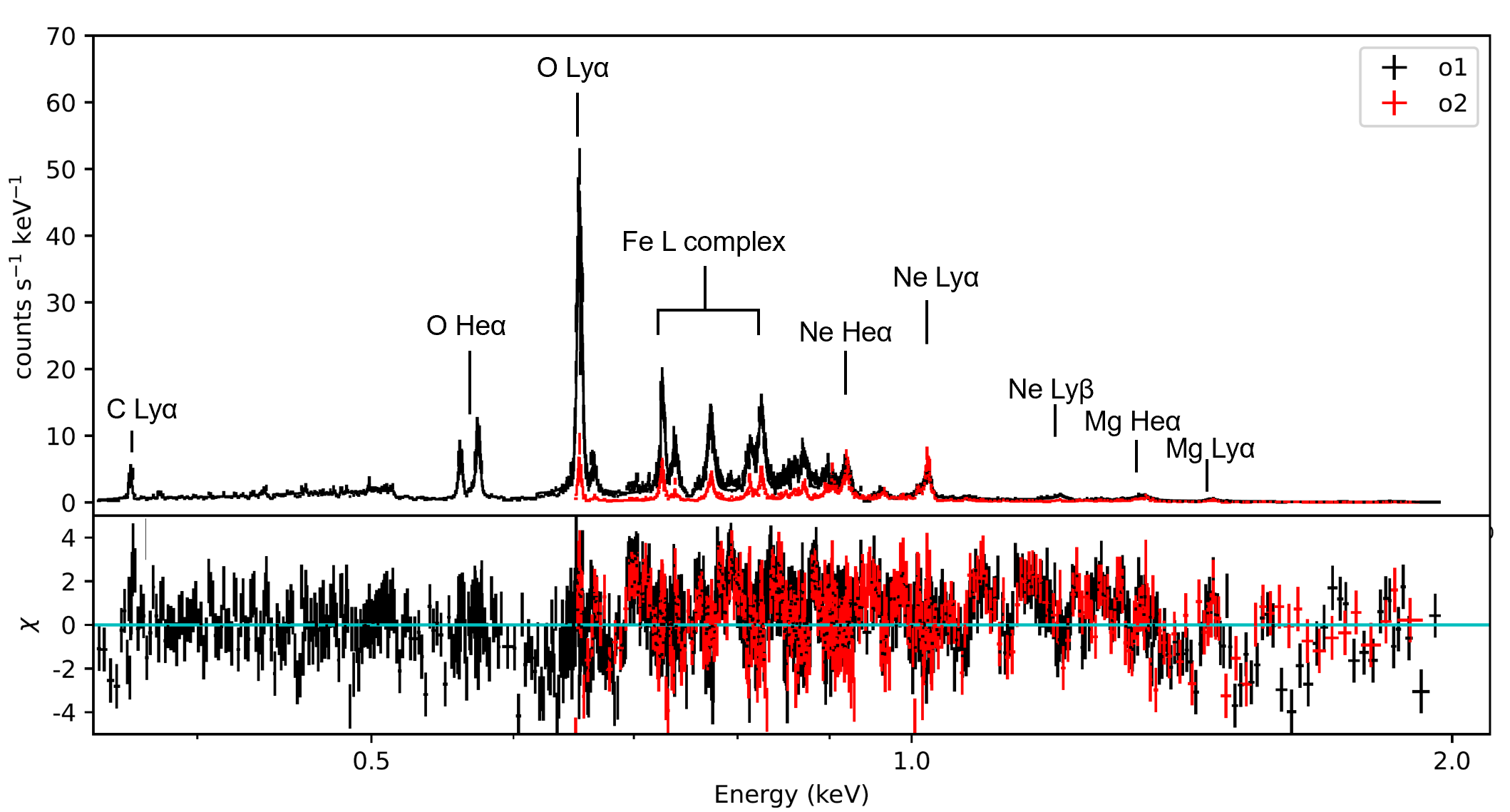}
    \caption{RGS spectra of N63A, with bright emission lines labeled.}
    \label{fig:rgs_line}
\end{figure*}

The high spectral resolution of the RGS allows us to resolve individual emission lines in the soft X-ray band. We focused on the 0.35--2.0\,keV range, where prominent lines from elements such as C, O, Ne, Mg, and Fe are clearly detected (Figure \ref{fig:rgs_line}).

In order to discriminate these emission lines and to obtain their fluxes, we used a thermal continuum component (\texttt{nlapec} in XSPEC) and some Gaussian line profiles (\texttt{Gauss} in XSPEC), where \texttt{nlapec} includes thermal bremsstrahlung, radiative recombination continua, and two-photon emission from collisionally-ionized diffuse gas. Both the continuum and line emission are subject to foreground absorption from the Milky Way (\texttt{TBabs} in XSPEC) and the LMC (\texttt{TBvarabs} in XSPEC). 
For the LMC absorption, we set the relative abundances to the average LMC values from \citet{Russell1992}, and fixed the two column densities to $N_{\rm{H,Gal}} = 6\times 10^{20}\,\rm{cm^{-2}}$, $N_{\rm{H,LMC}} = 5\times 10^{20}\,\rm{cm^{-2}}$ \citep{Dickey1990,Sano2019}, the latter being a typical value from previous studies of N63A.
We also included a Gaussian smoothing component (\texttt{gsmooth} in XSPEC) to account for intrinsic line broadening, convolving the spectrum with a Gaussian model whose width varies with energy as $\Sigma(E) = \sigma(E/6)^{\alpha}$, where $\sigma$ is the Gaussian width at 6\,keV and $\alpha$ controls the energy dependence.
A  
velocity component (\texttt{vashift} in XSPEC) was also included to model the line shift caused by the overall motion of the remnant.

\begin{figure*}
    \centering
    \includegraphics[width=0.75\linewidth]{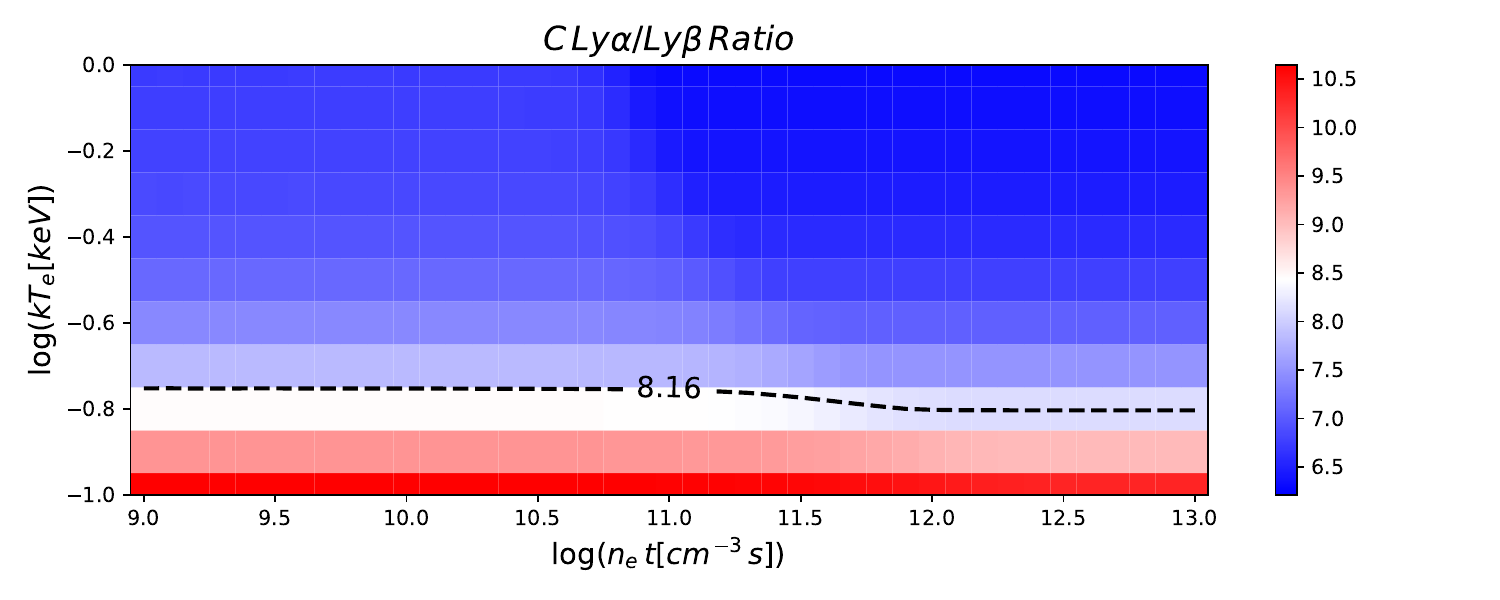}
    \caption{Theoretically predicted flux ratio of C Ly$\alpha$ to C Ly$\beta$. This figure is based on calculations using the SPEX code.}
    \label{fig:ratio}
\end{figure*}

Using this approach, we identified 38 emission lines in the 0.35--2.0\,keV RGS spectra, and listed their centroid energies and fluxes in  Table \ref{tab:gauss}. Below 0.5\,keV, the spectrum is dominated by H-like lines of C, while the 0.5--0.7\,keV range contains the He-like and H-like triplets of O, which are useful for diagnosing the plasma state. 
The clear detection of the C VI lines allows us to calculate the Ly$\alpha$ to Ly$\beta$ flux ratio, which is sensitive to temperature.  
The calculation is based on SPEX \citep{Kaastra1996} code and the theoretical predictions vary with temperature and $n_e t$ are shown in Figure \ref{fig:ratio}. The observed C Ly$\alpha$/Ly$\beta$ ratio in N63A is $8.16\pm2.49$. Comparing this value with the model  in Figure \ref{fig:ratio}, we find that the 8.16 contour lies within the vertical range of approximately -0.8 to -0.75 in the image, which corresponds to a temperature of about 0.16--0.18\,keV.
However, given the approximately 30\% uncertainty in this ratio, the corresponding temperature range in the simulated image is quite broad, encompassing multiple possible values. Consequently, the ratio derived from the best-fit value likely corresponds only to the most probable temperature range within the probability distribution. 
The best-fit velocity indicates a blueshift. Its magnitude is comparable to, but not fully consistent with the systemic blueshift of the LMC, which may be due to the difficulty in accurately determining the intrinsic line centroid and broadening for an extended source.

\begin{table}
     \centering
     \renewcommand\arraystretch{1.2}
     \setlength{\tabcolsep}{2pt}
	\caption{Identified Lines in RGS Spectra}
    \label{tab:gauss}
    \begin{tabular}{ccc}
         \hline
         Line & Centroid (keV) & Flux ($10^{-4}\,\rm{photons\,cm^{-2}\,s^{-1}}$)  \\
        \hline
         C Ly$\alpha$ & 0.3675 & $20.23_{-2.13}^{+2.14}$ \\
         S XIV & 0.3810 & $4.96_{-1.36}^{+1.37}$ \\
         C Ly$\beta$ & 0.4356 & $2.48_{-0.71}^{+0.71}$ \\
         C Ly$\gamma$ & 0.4594 & $0.93_{-0.52}^{+0.52}$ \\
         N Ly$\alpha$ & 0.5002 & $3.37_{-0.61}^{+0.62}$ \\
         O He$\alpha$-$f$ & 0.5611 & $31.87_{-1.68}^{+1.68}$ \\
         O He$\alpha$-$i$ & 0.5686 & $4.63_{-1.10}^{+1.09}$ \\
         O He$\alpha$-$r$ & 0.5740 & $5.10_{-1.93}^{+1.94}$ \\
         O Ly$\alpha$ & 0.6537 & $95.37_{-1.39}^{+1.39}$ \\
         O He$\beta$ & 0.6656 & $5.31_{-0.51}^{+0.51}$ \\
         Fe XVII & 0.7252 & $10.23_{-1.67}^{+1.67}$ \\
         Fe XVII & 0.7271 & $27.43_{-1.63}^{+1.63}$ \\
         Fe XVII & 0.7389 & $13.84_{-0.56}^{+0.57}$ \\
         Fe XVIII & 0.7715 & $8.11_{-0.88}^{+0.88}$ \\
         O Ly$\beta$ & 0.7746 & $22.38_{-0.96}^{+0.97}$ \\
         Fe XVII & 0.8124 & $12.00_{-0.69}^{+0.68}$ \\
         O Ly$\gamma$ & 0.8170 & $6.83_{-0.73}^{+0.73}$ \\
         Fe XVII & 0.8258 & $28.09_{-0.66}^{+0.66}$ \\
         Fe XVIII & 0.8531 & $7.66_{-0.43}^{+0.44}$ \\
         Fe XVIII & 0.8626 & $7.75_{-0.46}^{+0.46}$ \\
         Fe XVIII & 0.8726 & $14.50_{-0.51}^{+0.51}$ \\
         Fe XVII & 0.8968 & $7.27_{-0.47}^{+0.47}$ \\
         Ne He$\alpha$-$f$ & 0.9051 & $12.66_{-0.69}^{+0.69}$ \\
         Fe XIX & 0.9172 & $8.14_{-1.02}^{+1.01}$ \\
         Ne He$\alpha$-$r$ & 0.9220 & $24.22_{-1.05}^{+1.05}$ \\
         Fe XX & 0.9652 & $7.83_{-0.43}^{+0.43}$ \\
         Fe XXI & 1.0004 & $2.84_{-0.47}^{+0.48}$ \\
         Fe XXI & 1.0093 & $3.19_{-0.61}^{0.60}$ \\
         Ne Ly$\alpha$ & 1.0220 & $25.98_{-0.71}^{+0.71}$ \\
         Ne He$\beta$ & 1.0740 & $5.18_{-0.44}^{+0.44}$ \\
         Fe XXII & 1.0534 & $2.05_{-0.38}^{+0.38}$ \\
         Fe XXIII & 1.1291 & $2.36_{-0.32}^{+0.32}$ \\
         Ne Ly$\beta$ & 1.2110 & $3.58_{-0.28}^{+0.28}$ \\
         Mg He$\alpha$-$f$ & 1.3311 & $2.54_{-0.28}^{+0.29}$ \\
         Mg He$\alpha$-$r$ & 1.3523 & $4.96_{-0.34}^{+0.34}$ \\
         Mg Ly$\alpha$ & 1.4726 & $2.13_{-0.25}^{+0.25}$ \\
         Si He$\alpha$-$f$ & 1.8395 & $1.17_{-0.46}^{+0.45}$ \\
         Si He$\alpha$-$r$ & 1.8650 & $1.96_{-0.48}^{+0.48}$ \\ 
        \hline
        $\rm Velocity$ (km s$^{-1}$) & ... & $316_{-12}^{+12}$ \\
        $\sigma$ (eV) & ... & $0.11_{-0.02}^{+0.02}$ \\
        $\alpha$ & ... & $1.9_{-0.1}^{+0.1}$ \\
        \hline
    \end{tabular}
\end{table}

\subsection{Analysis on the Integral High-resolution Spectra}\label{Sec:xmm}
\subsubsection{Three-temperature (3-T) Component Model}\label{Sec:3T}
We jointly fitted the RGS (0.3--2.0\,keV) and EPIC (0.3--8.0\,keV) spectra of N63A with multi-temperature components (\texttt{vnei} in XSPEC) comprising 
several non-equilibrium ionization (NEI) components, all subjected to absorption, line broadening, and velocity.  
The  
NEI plasma model {\tt vnei} assumes a constant temperature and a single ionization timescale 
incorporates variable elemental abundances relative to solar values. The \texttt{gsmooth} and \texttt{vashift} components account for line broadening and systemic velocity shift, respectively.  
The model in XSPEC terminology is: 
\texttt{$\rm tbabs_{Gal}\times tbvarabs_{LMC}\times gsmooth\times vashift\times (N\times vnei)$}.

The Galactic absorption column density ($N_{\rm{H,Gal}}$) was fixed at $6\times10^{20}\,\rm{cm^{-2}}$ \citep{Dickey1990}, while the LMC absorption column ($N_{\rm{H,LMC}}$) was left free, with its elemental abundances fixed to the average LMC values from \citet{Russell1992}. 
For the plasma components, the abundances of the low-temperature NEI component were fixed to the LMC abundances. For the other two components, the abundances of O, Ne, Mg,  Si, S, and Fe were linked together and set as free parameters, while all other elements were fixed to LMC values, e.g., He (0.89), C (0.46), N (0.18), Ca (0.49), Ni (0.98). 
Initial fitting with a simpler two-temperature model revealed significant positive residuals above $\sim 4$\,keV, indicating an excess of high-energy emission. 
To explain this, we explored two physical scenarios by extending the model with either an additional high-temperature thermal plasma component or a non-thermal power-law component. 
The addition of a third NEI component led to a substantial improvement in the fit (reduced $\chi^2$=1.42), effectively accounting for the high-energy residuals. 
To quantify the contribution of the background at high energies, 
we performed a fit to the background spectrum, and attempted to include the best-fit background model as a component in the fitting model for the source spectrum and perform the fit. 
The details on background spectral fitting are presented in Appendix B.
Over the 4--8\,keV band, the total flux of the source is approximately $6.3\times10^{-13}\,\rm erg\,cm^{-2}\,s^{-1}$, whereas the background contribution accounts for about 5\%, with a flux of roughly $2.9\times10^{-14}\,\rm erg\,cm^{-2}\,s^{-1}$. For consistency with the \textit{Chandra} spatially resolved spectral analysis described later, however, we retain the spectral results derived from direct background subtraction. 

Alternatively, the model with a power-law component also reduced the high-energy residuals, yielding a photon index of $\Gamma \approx$2.80$\pm$0.06. 
However, its fit quality was 
(reduced $\chi^2$=1.51) not as good as that of the 3VNEI model. 
To establish a more robust statistical justification for the model selection, we employed the Akaike Information Criterion (AIC). The AIC provides a quantitative measure for comparing statistical models by balancing goodness of fit against model complexity, thereby penalizing unnecessary free parameters that may lead to overfitting. For spectral fitting performed in XSPEC, where the fit statistic is evaluated using the $\chi^2$ statistic, the AIC is defined as $\rm{AIC} = \chi^2 + 2$$k$, where $\chi^2$ is the value of the fit statistic obtained from the spectral fitting, and $k$ denotes the number of free parameters in the model. The difference between two models is used to assess the relative support for each model. A value of $\Delta \rm AIC>10$ is generally considered strong evidence in favor of the model with the smaller AIC \citep{Burnham2002}. 
The calculated AIC values for the 
three competing models are as follows: 
for the 2VNEI model, AIC=4695.94+20=4715.94;
for the 3VNEI model, AIC=3857.87+23=3880.87; for the two-temperature VNEI plus power-law (2VNEI+PowerLaw) model, AIC=3965.17+22=3987.17. The resulting difference of first two models is $\Delta_1 \rm AIC=835.1$, the difference of last two models is $\Delta_2 \rm AIC=106.3$, which substantially exceeds the threshold of 10. This large $\Delta \rm AIC$ value provides decisive statistical support for the 3VNEI model, indicating that the addition of a third thermal component is favored over the inclusion of a non-thermal power-law tail for describing the high-energy excess in the spectrum of N63A.
Considering both the statistical superiority and the physical plausibility of very hot plasma in an SNR, our discussion is mainly based on the three-temperature model.  
We ran a Monte Carlo Markov chain (MCMC), using the XSPEC command \texttt{chain}, in order to get 1$\sigma$ error. 

\begin{figure}
    \centering
    \includegraphics[width=\linewidth]{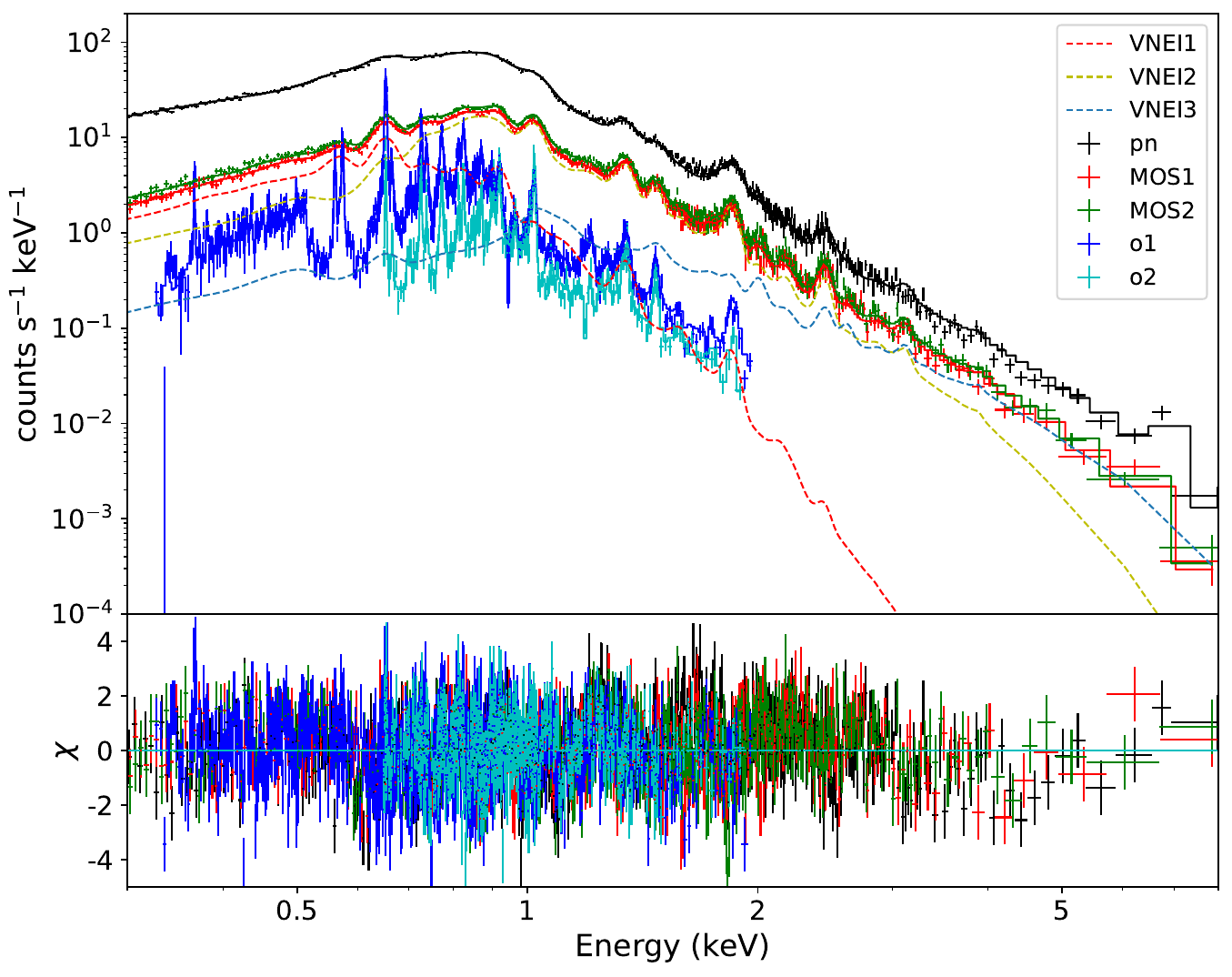}
    \caption{Global XMM-\textit{Newton} spectra of N63A fitted with the triple-temperature model.}
    \label{fig:XMMspec}
\end{figure}

The best-fit spectra and residuals are shown in Figure \ref{fig:XMMspec}, with parameters listed in Table \ref{tab:3T}. 
The MCMC results and related descriptions are presented in Appendix C.

The three 
NEI components 
are at  
temperatures 
$kT\sim$0.3, 0.7 and 1.5\,keV, respectively.
The volume emission measure (VEM) is dominated by the low-temperature component, followed by the medium-temperature one, while the highest-temperature component contributes only a small fraction. The ionization timescale $n_et$ shows a slightly positive correlation with temperature, where $n_e$ is the electron density, and $t$ is the time elapsed since the gas was shocked. 
The abundances of O, Ne, Mg, Si and S are all about twice the average level in the LMC, while Fe is very close to the ISM level.

\begin{table}
     \centering
     \renewcommand\arraystretch{1.4}
     \setlength{\tabcolsep}{4pt }
	\caption{Spectral Fitting Results of 3-T Models}
    \label{tab:3T}
    \begin{tabular*}{0.5\textwidth}{@{\extracolsep{\fill}}cccl@{}}
         \hline\hline
         Component & Parameter & \\
         \hline
         TBvarabs & $N_{\rm H,LMC}(10^{20}\,\rm cm^{-2})$ & $7.79_{-0.27}^{+0.25}$ \\
         vashift & Velocity ($\rm km s^{-1}$) & $206.6_{-0.5}^{+0.6}$ \\
         gsmooth & $\sigma$ (eV) & $3.3_{-0.5}^{+0.5}$ \\
                & $\alpha$ & $0.88_{-0.06}^{+0.10}$ \\
        VNEI1 & $kT_{e,1}$ (eV) & $318_{-9}^{+11}$ \\
          & $n_e t$ ($10^{11}\,\rm cm^{-3}\,s$) & $1.15_{-0.10}^{+0.08}$ \\
          & norm\tnote{a} ($10^{-2}\,\rm cm^{-5}$) & $3.46_{-0.18}^{+0.14}$ \\
        VNEI2 & $kT_{e,2}$ (eV) & $718_{-3}^{+3}$ \\
          & $n_e t$ ($10^{11}\,\rm cm^{-3}\,s$) & $6.68_{-0.48}^{+0.53}$ \\
          & norm\tnote{a} ($10^{-2}\,\rm cm^{-5}$) & $3.98_{-0.13}^{+0.14}$ \\
        VNEI3 & $kT_{e,3}$ (eV) & $1481_{-28}^{+37}$ \\
          & $n_e t$ ($10^{11}\,\rm cm^{-3}\,s$) & $9.43_{-2.23}^{+3.48}$ \\
          & norm\tnote{a} ($10^{-3}\,\rm cm^{-5}$) & $9.22_{-0.47}^{+0.39}$ \\
          & O ($Z/Z_{\odot}$)  & $1.12_{-0.07}^{+0.07}$ \\
          & Ne ($Z/Z_{\odot}$)  & $1.64_{-0.10}^{+0.10}$ \\
          & Mg ($Z/Z_{\odot}$)  & $0.94_{-0.04}^{+0.04}$ \\
          & Si ($Z/Z_{\odot}$)  & $0.98_{-0.04}^{+0.04}$ \\
          & S ($Z/Z_{\odot}$)  & $0.80_{-0.04}^{+0.04}$ \\
          & Fe ($Z/Z_{\odot}$)  & $0.44_{-0.02}^{+0.02}$ \\
          
          \hline
           & reduced-$\chi^2$ (d.o.f) & 1.42(2694) \\

         \hline
    \end{tabular*}

\end{table}    

\subsubsection{DEM Model}
To better understand the plasma conditions and temperature distribution in N63A, we also applied a 
differential emission measure (DEM) analysis for comparison.
The DEM model employed in this study 
has been described by \citet{Sun2025}, which is basically a more comprehensive version of XSPEC's \texttt{c6pvmkl} model. In this framework, the DEM function is derived from the 
VEM of thermal plasma, describing the temperature distribution of the plasma, and is parameterized by a 7$^{th}$ order Chebyshev polynomial. The temperature range of 0.1-10\,keV is logarithmically divided into 40 bins, with each bin modeled by an independent \texttt{vnei} component. The ionization timescale is simplified as a power-law function of temperature:
\begin{equation}
    \tau_i = \tau_{1\,\rm{keV}}\left(\frac{T_i}{1\,\rm{keV}}\right)^\beta,
\end{equation}
where $\tau_i = (n_et)_i$, 
$i$ represents a specific temperature bin. $\tau_{1\,\rm{keV}}$ is the ionization parameter at 1\,keV and $\beta$ is the power-law index.

\begin{figure}
    \centering
    \includegraphics[width=0.95\linewidth]{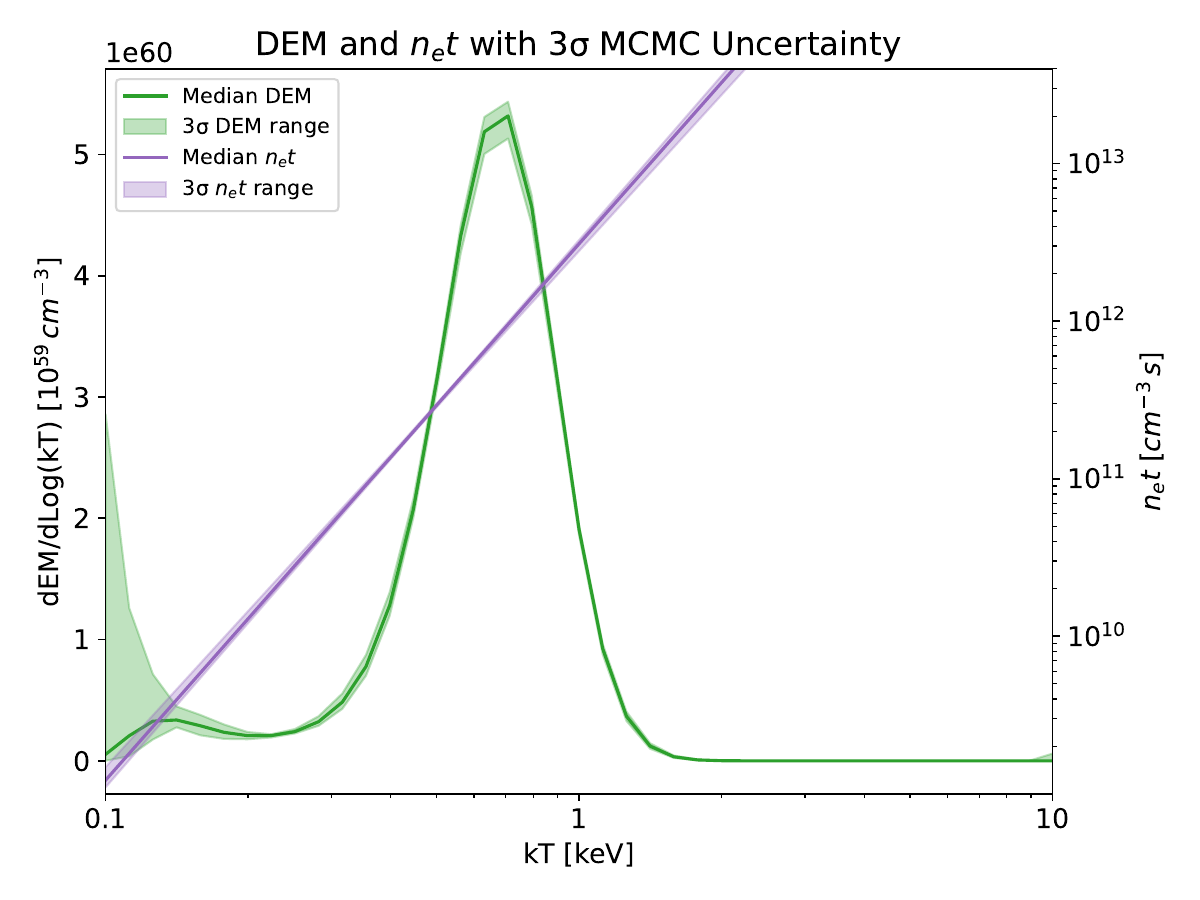}
    \caption{DEM fitting results for N63A based on XMM-\textit{Newton} RGS and EPIC observations.}
    \label{fig:dem}
\end{figure}

We constrained the DEM distribution and its parameters by fitting the global RGS and EPIC spectra. The spectral model consists of a specific DEM model named \texttt{c7pvnei}, two foreground absorption components ($\rm tbabs_{Gal}$ and $\rm tbvarabs_{LMC}$), a velocity component and a Gaussian smoothing component 
with the same settings as 3-T fitting (Section \ref{Sec:3T}). To summarize, this model can be described as \texttt{$\rm{tbabs_{Gal}\times tbvarabs_{LMC}\times gsmooth\times vashift\times c7pvnei }$}. After the spectral fitting, 
we also ran a MCMC with the same settings as described in the previous section. The result of the temperature distribution is represented in 
Figure \ref{fig:dem}.
The dominant temperature distribution lies in the range of 0.4--1.0\,keV, with the VEM exhibiting a peak around 0.7\,keV. An excess is also present below 0.2\,keV, although it is poorly constrained. The ionization timescale shows a positive correlation with the temperature. 

However, the simplified setup of this model, particularly the tied abundances of 40 temperature bins and the assumed power-law relation for $n_e t$, likely leads to unphysical behavior at the low- and high-temperature extremes of the reconstructed DEM distribution. Specifically, the model struggles to account for the high-energy emission above $\sim 4$\,keV. However, when we introduced a non-thermal power-law component to compensate for this deficiency, the fit yielded a very soft photon index ($\Gamma\sim 3.15$). Such a soft spectrum is uncharacteristic of typical particle acceleration in SNRs, casting doubt on the physical reality of this component and leaving the origin of the hard excess ambiguous within the DEM framework.

Given these limitations and the fact that the DEM model did not provide a statistically superior fit compared to the 3VNEI model, we regard its results as less reliable for drawing firm conclusions about the highest-temperature plasma or potential non-thermal emission. Therefore, we base our primary interpretation on the more robust 3VNEI model and use the DEM analysis only as a supplementary view of the general temperature structure, whose peak around 0.7\,keV remains a consistent and credible feature.

\subsection{Spatially Resolved Spectral Analysis}\label{sec:chandra}
\textit{Chandra} data provide higher angular resolution and longer exposure times, enabling us to resolve the fine structural details of N63A. This makes it highly suitable for cross-calibration with our XMM-\textit{Newton} fitting results. We therefore performed a spatially resolved spectroscopic analysis across the entire remnant.

To ensure a sufficiently high signal-to-noise ratio (S/N) in each extraction region, we applied the WVT binning algorithm \citep{Diehl2006}, which is a generalization of \citet{Cappellari2003} Voronoi binning algorithm. 
A 0.3–-7.0\,keV broadband counts image of an elliptical region 
covering N63A was generated, and the WVT algorithm divided the entire area into 106 bins, each containing approximately 10,000 counts (S/N $\sim$ 100). Spectra from  
these bins were then extracted and fitted. 
All of the spectra were rebinned to 
include at least 10 counts per spectral bin. 
The background spectrum was extracted from a 1.5 arcmin-wide source-free annular 
and surrounding the remnant.

The highest-temperature component in the 3-T model becomes dominant only at energies $\geq$4\,keV. However, due to the limited counts in the \textit{Chandra} spectrum, the number of high-energy photons is very low, which prevents this hottest component from being well constrained. 
This discrepancy is primarily attributed to the substantial difference in the effective area between the two instruments. While \textit{Chandra} provides superior spatial resolution, its smaller effective area leads to 
insufficient photon statistics and 
S/N for the faint hard X-ray emission. Conversely, the high throughput of XMM-\textit{Newton} is critical for robustly constraining this high-temperature plasma component.
Therefore, we fitted the spectrum from each region 
with a model consisting of  
two NEI components (\texttt{vnei}), subject to the same Galactic and LMC absorption components as described in Section \ref{Sec:3T}.
The elemental abundances of the  
first NEI component were fixed to the average abundance of LMC ISM.
For the second NEI component, the abundances of O, Ne, Mg, Si, S and Fe were set as free parameters.

\begin{figure*}
    \centering
    \includegraphics[width=1.0\linewidth]{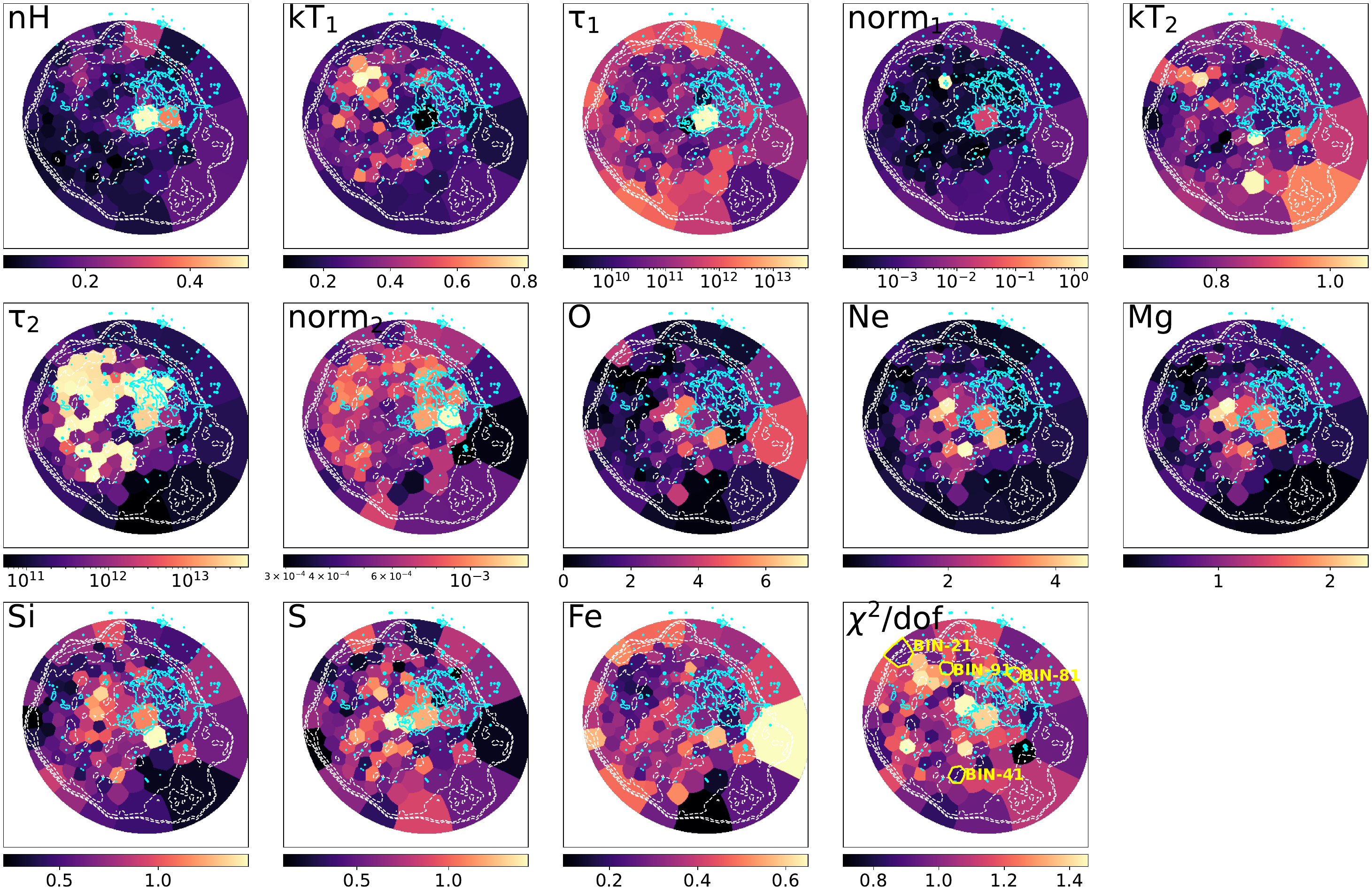}
    \caption{Maps of individual spectral fitting parameters: column density ($10^{22}\,\rm cm^{-2}$), electron temperature $kT$ (keV), ionization timescale $\tau$ ($\rm s\,cm^{-3}$), normalization ($\rm cm^{-5}$), metal abundances and reduced-$\chi^2$. The white contours represent the X-ray emission, while the cyan contours outline the optical nebula. The four yellow polygons denote the regions used to extract the example spectra in Table \ref{tab:wvtspec} and Figure \ref{fig:wvt_spec}.}
    \label{fig:wvt}
\end{figure*}

The resulting parameter maps are shown in Figure \ref{fig:wvt}, and selected examples of the spectral fits are provided in Table \ref{tab:wvtspec}. The first temperature component is centered around 0.2--0.3\,keV, while the second is around 0.7--0.8\,keV. The ionization timescale of the first NEI component is generally lower than that of the second. Additionally, 
the metal abundances are found to be significantly elevated in the high-temperature component. All these findings are consistent with the results from XMM-\textit{Newton}. The temperature of the  
low-temperature component is higher in the outer regions of the remnant and lower in the central areas, particularly near the optical nebulae. The regions coinciding with the optical nebula exhibit higher absorption and  
emission measure. Notably, the two eastern lobes of the optical nebula show significantly higher ionization timescales compared to the western lobe. This spatial dichotomy in the ionization state aligns 
well with the optical classification, where the eastern lobes are confirmed to be shock-ionized while the western lobe is predominantly photoionized \citep{Levenson1995}.

The elemental abundances provide crucial clues for understanding the SN nucleosynthesis. In N63A, O, Ne and Mg show higher enhancements compared to Si, S and Fe. Spatially, the abundances of O, Ne, Mg, Si, and S are elevated in the central regions of the remnant, with several prominent overabundance peaks, suggesting an asymmetric distribution of the mixed ejecta. Notably, enhanced Fe abundances are observed in the outermost regions, which may indicate the presence of high-velocity, Fe-rich ejecta clumps.

\begin{table*}
     \centering
     \renewcommand\arraystretch{1.4}
     \setlength{\tabcolsep}{4pt }
	\caption{Examples of \textit{Chandra} Spectral Results}
    \label{tab:wvtspec}
    \begin{tabular*}{0.7\textwidth}{@{\extracolsep{\fill}}ccccccl@{}}
         \hline\hline
        Component & Parameter & BIN-21 & BIN-41 & BIN-81 & BIN-91  \\
         \hline
         TBvarabs & $N_{\rm H,LMC}(10^{20}\,\rm cm^{-2})$ & $10.00_{-1.13}^{+4.24}$  & $7.65_{-1.67}^{+7.34}$ & $8.71_{1.17}^{+1.34}$ & $13.62_{-0.50}^{+0.42}$\\
         VNEI1 & $kT_{e,1}$ (keV) & $0.23_{-0.03}^{+0.06}$ & $0.27_{-0.02}^{+0.02}$ & $0.37_{-0.04}^{+0.05}$ & $0.33_{-0.01}^{+0.01}$\\
          & $n_e t$ ($10^{11}\,\rm cm^{-3}\,s$) & $2.08_{-1.15}^{+10.65}$  & $1.89_{-0.64}^{+1.62}$ & $0.82_{-0.27}^{+0.45}$ & $2.34_{-0.30}^{+0.43}$\\
          & norm\tnote{a} ($10^{-3}\,\rm cm^{-5}$) & $1.19_{-0.39}^{+1.46}$  & $0.65_{-0.13}^{+0.27}$ & $0.43_{-0.07}^{+0.10}$ & $0.57_{-0.03}^{+0.03}$\\
         VNEI2 & $kT_{e,2}$ (keV) & $0.75_{-0.03}^{+0.03}$ & $0.76_{-0.01}^{+0.02}$ & $0.86_{-0.02}^{+0.03}$ & $0.76_{-0.02}^{+0.36}$\\
          & $n_e t$ ($10^{11}\,\rm cm^{-3}\,s$) & $4.50_{-1.71}^{+5.26}$  & $7.98_{-1.17}^{+2.49}$ & $3.04_{-0.93}^{+1.75}$ & $5.32_{-0.83}^{+1.45}$ \\
          & norm\tnote{a} ($10^{-4}\,\rm cm^{-5}$) & $4.39_{-1.29}^{+1.36}$  & $4.47_{-0.57}^{+0.56}$ & $5.18_{-0.20}^{+0.24}$ & $8.40_{-0.26}^{+0.28}$\\
          & O ($Z/Z_{\odot}$) & $3.82_{-1.72}^{+3.18}$  & $1.74_{-0.43}^{+0.56}$ & $1.36_{-0.33}^{+0.26}$ & $0.35_{-0.13}^{+0.10}$\\
          & Ne ($Z/Z_{\odot}$) & $0.94_{-0.54}^{+0.95}$  & $2.33_{-0.73}^{+0.60}$ & $0.42_{-0.19}^{0.36}$ & $0.51_{-0.09}^{+0.07}$\\
          & Mg ($Z/Z_{\odot}$) & $0.59_{-0.21}^{+0.35}$  & $1.06_{-0.24}^{+0.46}$ & $0.60_{-0.11}^{+0.13}$ & $0.45_{-0.05}^{+0.05}$\\
          & Si ($Z/Z_{\odot}$) & $0.72_{-0.20}^{+0.35}$  & $1.18_{-0.22}^{+0.28}$ & $0.57_{-0.10}^{+0.11}$ & $0.62_{-0.07}^{+0.08}$\\
          & S ($Z/Z_{\odot}$) & $0.56_{-0.28}^{+0.37}$  & $0.66_{-0.29}^{+0.28}$ & $0.11_{-0.11}^{+0.16}$ & $0.93_{-0.20}^{+0.19}$\\
          & Fe ($Z/Z_{\odot}$) & $0.52_{-0.13}^{+0.25}$  & $0.42_{-0.06}^{+0.08}$ & $0.36_{-0.02}^{+0.02}$ & $0.26_{-0.01}^{+0.02}$\\
          \hline
          & reduced-$\chi^2$(d.o.f) & 0.93 (113) & 0.86 (109) & 1.14 (111) & 0.96 (118) \\
          \hline

    \end{tabular*}

\end{table*}

\section{Discussion}\label{sec:discuss}
\subsection{Properties of N63A}\label{properties}
We derived the parameters of the X-ray emitting plasma based on the spectral fitting results from both XMM-\textit{Newton} and \textit{Chandra}. The hydrogen density $n_{\rm{H}}$ was obtained from the normalization parameter, which is defined as:
\begin{equation}
    norm = \frac{10^{-14}}{4\pi d^2}\int n_{\rm{e}}n_{\rm{H}}f\rm{dV}
\end{equation}
where $d$=50\,kpc is the distance to the LMC, $n_{\rm{e}}$ is the electron density ($n_{\rm{e}} = 1.2 n_{\rm{H}}$ for a fully ionized plasma with solar abundance), $f$ is the volume filling factor, and $V$ is the volume of the source. 
For XMM-\textit{Newton} results, we assumed that the 
three components (designated as components 1, 2 and 3, with $kT\sim 0.3$\,keV, 0.7\,keV and 1.5\,keV, respectively see Table \ref{tab:3T}) share the same volume (with filling factors $f_1$, $f_2$ and $f_3$, respectively), 
i.e., $f_1+f_2+f_3 = f\leq 1$.
To estimate the volume filling factors and densities of the three temperature components, we adopted a simplified pressure-balance framework among the components, where $n_{\rm{e,1}}kT_{\rm{e,1}} = n_{\rm{e,2}}kT_{\rm{e,2}}=n_{\rm{e,3}}kT_{\rm{e,3}}/6$. The physical motivation of this framework is discussed in Section \ref{sec:temperature}. 
The volume $V$ for the three components  
was taken as the volume of a sphere with radius $R=9$\,pc. 
Thus, we obtained the filling factors 
$f_1=0.142 \pm 0.005 f$, $f_2=0.835 \pm 0.006 f$ and $f_3=0.023 \pm 0.001 f$
hydrogen number densities  
$n_{\rm H,1}=8.20 \pm 0.3f^{-1/2}\,\rm cm^{-3}$, $n_{\rm H,2}=3.63 \pm 0.05f^{-1/2}\,\rm cm^{-3}$ and $n_{\rm H,3}=10.55 \pm 0.23 f^{-1/2}\,\rm cm^{-3}$ for the three components. Taking $M=1.4m_H n_{\rm H} V$, the masses of the X-ray emitting gas are derived as $M_1= 123 \pm 3 f^{1/2}\,M_{\odot}$, $M_2= 320 \pm 6 f^{1/2}\,M_{\odot}$ and $M_3= 25 \pm 1  f^{1/2}\,M_{\odot}$. The large derived filling factor and mass for component 2, comparable to that of component 1, indicate that it is predominantly composed of shocked medium rather than pure ejecta. This is consistent with the lack of strong abundance enhancements in the global fit (Table \ref{tab:3T}). 
We interpret this component as metal enriched interclump medium.
Overall, these results suggest a scenario in which the SNR evolves in a clumpy ambient medium, where the intermediate-temperature component dominates the volume and mass and thus primarily traces the forward-shocked ambient gas.

We used the properties of component 2 to estimate the dynamical properties of the remnant. The velocity of the forward shock  
is estimated as $v_s = [16kT/(3\mu m_{\rm H})]^{1/2} 
\sim 780\,\rm km\,s^{-1}$, where $\mu=0.61$ is the mean atomic weight per hydrogen atom for a fully ionized plasma, and $m_{\rm H}$ is the mass of a hydrogen atom. The Sedov age is given by $t_{\rm Sedov} = 2R_s/5v_s \sim 4.5\,kyr $. The explosion energy can be obtained from $E_0 = 25(1.4n_0 m_{\rm H})R_s^3 v_s^2/(4\xi)\sim1.03\times10^{51}n_0\,\rm{erg}$, where $\xi=2.026$.  
We took $n_0 \sim n_{\rm H,2}$ and $f\sim 1$, as the intermediate-temperature component dominates the X-ray–emitting mass and volume and is expected to be representative of the swept-up ambient medium, rather than applying the ideal strong-shock compression factor of 4. The explosion energy is approximately $4\times10^{51}\,\rm{erg}$. 
High explosion energy $>1\times10^{51}$\,erg is also seen in other LMC SNRs, such as N132D \citep[e.g., ][]{Chen2003,Bamba2018}, N49B \citep[e.g., ][]{Hughes1998}. 
In the LMC, the relatively low metallicity (0.3--0.5 solar) may favor high explosion energies. Reduced line-driven mass loss allows massive stars to retain more massive cores and angular momentum prior to collapse, potentially leading to more energetic explosions compared to their solar-metallicity counterparts \citep{Meynet2005,Heger2003}. 
The exceptionally small formal errors returned by the MCMC analysis reflect the high statistical quality of the spectral data and the tight parameter constraints achieved with the 3-T model, indicating that the statistical uncertainty is very small. These uncertainties, however, neglect intrinsic parameter correlations and, more importantly, the systematic errors introduced by the idealized Sedov assumption of a uniform ambient medium, as well as by data calibration uncertainties, uncertainties in the spectral model and atomic data, and other physical model assumptions. Consequently, systematic uncertainty dominates the overall error budget. We therefore treated these values as rough estimates, but they are still accurate enough for the purposes of our study.

Similarly, calculations were performed for the 
spatially-resolved small-scale analysis of the \textit{Chandra} observation.
For computational simplicity, we omitted the outermost ring of the WVT regions, as it includes some background emission and primarily samples the faint ear-like structures.
The volume of each region was approximated as a cylinder, with a line-of-sight depth $l$ determined by its position relative to the SNR center and the assumed shell geometry. 
For each region, $l = 2\sqrt{R^2 - r^2}$
, where $r$ is the distance from the center of the region to the SNR center.
The area S of each region was determined from the number of pixels and the pixel size, giving the volume $V = S \times l$.
Assuming pressure balance and a common volume for the 
two components in the \textit{Chandra} fits, we estimated the total mass of the X-ray emitting gas by individually calculating the density, volume, and mass for all small regions and summing them. The resulting mass estimates are approximately  
$M_1\sim 120\,M_{\odot}$ and $M_2\sim 350\,M_{\odot}$, which are  
similar to the estimates from the XMM-\textit{Newton} data analysis.

For comparison, we  
re-performed the rim region fit 
as those in \citet{Karagoz2023} and obtained similar results for temperature, elemental abundances, and ionization timescale. However, 
a discrepancy arises in the density estimate. In \citet{Karagoz2023}, 
the first component was interpreted as shocked ISM with abundances fixed to the LMC ISM values, and the pre-shock density was derived as $n_0 \sim 6\,\rm cm^{-3}$, based on an assumed line-of-sight depth of $\sim0.56$\,pc for the rim region. In our analysis, the hydrogen density of the low-temperature component is $n_{\rm H,1} \approx 10.6\,\rm cm^{-3}$, which is a factor of $\sim2.4$ smaller than the value of $4n_0$ reported by \citet{Karagoz2023}. We argue that this difference can be naturally explained by the volume assumption. The adopted line-of-sight depth of 0.56\,pc is likely too small to be physically representative of the rim geometry.
If the volume is instead estimated using a spherical crown or cylindrical geometry based on the extent of the rim, the resulting volume would be 2--3 times larger. Consequently, the derived pre-shock density $n_0$ would be lower. 
Importantly, this difference in the local rim density does not imply a lower explosion energy in our analysis. Our estimate of the explosion energy is based on the density of the dominant X-ray-emitting component, rather than on the rim density alone. Consequently, our revised energy estimate remains relatively high.

\subsection{Temperature Distribution}\label{sec:temperature}
Our global spectral analysis of the XMM-\textit{Newton} 
observation reveals three distinct thermal plasma components in N63A, with temperatures  
around 0.3\,keV, 0.7\,keV, and 1.5\,keV, respectively. The presence of multiple temperature components is a characteristic signature of a SNR expanding into a non-uniform, multi-phase ISM. The spatially-resolved spectral analysis  
of \textit{Chandra} observation, which offers  
relatively high spatial resolution, primarily identifies two temperature components, with the highest temperature component ($\sim$1.5\,keV) not being clearly detected. 

In the 2vnei+powerlaw case, the photon index ($\sim2.8$) obtained for the power-law component falls within the range typical of non-thermal emission from SNRs, so the presence of a non-thermal contribution cannot be completely ruled out. However, the RGB image of N63A (Figure \ref{fig:chandra_image}) shows that the hard X-ray emission is predominantly concentrated in the interior of the remnant, especially near the optical nebula, rather than at the outer edge where particles are accelerated by blast shock (such as in SNRs Cas A, Kepler, Tycho, etc.). This spatial distribution disfavors a non-thermal origin. Besides, the spectral fitting results and statistical calculations presented earlier in Section \ref{Sec:xmm} also prefer a high-temperature thermal component. We therefore focus on this thermal scenario in the following discussion of this section.

We interpret the origins of the three distinct thermal components within the framework of a blast wave  
propagating into an inhomogeneous environment containing both a tenuous inter-cloud medium and dense 
cloudlets. 
The low-temperature component likely originates from the evaporation of shocked dense clouds.
This scenario is 
supported by both morphological and quantitative constraints. 
High-resolution HST WFPC2 observations  
reveal numerous shocked cloudlets with characteristic sizes down to 0.1\,pc embedded within the remnant,  
demonstrating that N63A is expanding into a highly clumpy environment \citep{Chu1999}. Several of these cloudlets exhibit diffuse morphologies and radial density gradients, which \citet{Chu1999} interpreted as suggestive of isotropic evaporation. 
From the spectral modeling, this component is characterized by a relatively small volume filling factor ($\sim15\%$), while exhibiting one of the highest densities among the three thermal components. Such a combination of small filling factor, high density, and moderate mass fraction is a hallmark of evaporated medium originating from dense cloudlets rather than from the shocked ISM.
In addition, both the low- and intermediate-temperature components  
exhibit densities significantly higher than that of the typical ISM, consistent with strong shock-cloud interactions and subsequent mass loading of the hot plasma. The relatively low ionization timescale of the low-temperature component further suggests that this plasma has been heated to X-ray-emitting temperatures only recently, as expected if the cloudlets have been recently evaporated \citep[e.g.,][]{Chen2004}.

These observational properties are in good agreement with the numerical simulations of \citet{Zhang2019}, who investigated the evolution of SNR expanding into a cloudy ISM under NEI conditions. Their simulations predict the coexistence of a dominant blast-shocked inter-cloud medium (ICM) and a secondary, cooler, high-density component with a small volume filling factor, produced by the evaporation. The temperature contrast and density structure inferred in their models closely resemble the properties of the low-temperature component observed in N63A.

The intermediate-temperature component (i.e., component 2) primarily consists of the shocked ICM, mixed with a portion of the SN ejecta. This component has a lower density but a larger total mass 
than the low-temperature component, which is difficult to explain if it were pure ejecta.
The mild enhancements observed in the abundances of O, Ne, Mg, Si, and S relative to the average LMC ISM provide the chemical evidence for the incorporation of nucleosynthesized material from the ejecta.

The origin of the highest-temperature plasma can be attributed to multiple reflected shocks when
the SN blast wave interacts 
with dense clouds 
heating the gas to high temperatures \citep{Sano2019}. 
When the incident shock encounters a high-density cloud, a portion of its energy is reflected backward. In the region between the reflected wave and the clump/cloud surface, the pressure increases again, and the corresponding jumps in density and pressure depend on the Mach number of the blast wave. When the Mach number exceeds the critical value of $M_s=2.76$ \citep{Spitzer1982}, the reflected shock can form a stable bow shock in front of the clump/cloud. In the strong shock limit, the local pressure ratio reaches a peak value of $p_{\rm II}/p_{\rm I}\approx6$, which corresponds to a temperature elevation of 2.4 times ($T_{\rm II}/T_{\rm I} = 2.4$) and 2.5 times for density ($\rho_{\rm II}/\rho_{\rm I}=2.5$) \citep{Hester1986,Spitzer1982}
, where the subscripts  
I and II refer to the conditions behind the blast wave and behind the reflected shock, respectively. This predicted temperature and density contrast is broadly consistent with the observed ratio between the intermediate- and high-temperature components in N63A, suggesting that reflected shocks may play a key role in producing the hotter X-ray emitting plasma than the blast-shocked gas.

\subsection{Correlation Between $n_e t$ and $kT_e$}\label{sec:net}
The DEM fitting results 
show a power-law 
assumption between the ionization timescale and temperature, with the index ($\beta$) being a relatively large positive value. 
Notably, the 3-T fit based on XMM-\textit{Newton} and two-temperature fit based on \textit{Chandra} also show a positive correlation between $n_e t$ and $kT_e$, consistent with the DEM results, indicating that this correlation is not a fitting artifact introduced by the DEM method but originates from the physical information constrained in the observational data. However, the physical origin of this is not unique, and its interpretation must be distinguished by considering the physical sources of different temperature components.

In the previous discussion, we have indicated that the low- and high-temperature plasma in N63A do not originate from the same physical process, but are instead associated with cloud evaporation and the reflected shock, respectively. Within this context, the measured $n_e t$ values corresponding to different temperature components should be understood as the ionization evolution times elapsed since each component was heated to its X-ray emitting temperature. 
For the low-temperature component, its ionization timescale is notably low. Combined with its higher-density, small volume filling factor, and low mass fraction, this characteristic naturally points to a plasma component that has been recently formed. In the cloud evaporation scenario, after being 
engulfed by the blast shock, dense clouds do not immediately contribute to X-ray emission. Instead, they gradually evaporate via thermal conduction and mixing processes \citep{Zhang2019}. The ionization history effectively begins only after the medium enters the hot phase. Therefore, the lower $n_e t$ of the low-temperature component indicates that these cloudlets have only recently completed evaporation and been heated to X-ray-emitting temperatures. Based on the density estimates derived earlier from 
VEM and pressure equilibrium, we can constrain 
the order of magnitude of the timescale for this process. Taking the electron density of the low-temperature component and its measured $n_et$ value, the resulting characteristic time is on the order of $10^3$\,yr, which is significantly shorter than the dynamical age of N63A. This result supports the interpretation that the low-temperature plasma did not exist in the early stages of the remnant but is closely linked to ongoing cloud-shock interactions and subsequent evaporation processes.

In contrast, the high-temperature component exhibits a relatively high ionization timescale despite its small volume filling factor. In the reflected shock scenario, the high-temperature plasma is generated when the forward shock encounters a dense cloud, an event that likely precedes the significant evaporation phase of the cloud. Although this component is highly localized in space and involves a limited amount of masses, its early formation and subsequent ionization evolution over a longer period allow its $n_e t$ value to reach a high level. Thus, the characteristic high $n_e t$ of the high-temperature component is  consistent with the early generation and long-term evolution associated with the reflected shock.

The intermediate-temperature component occupies an intermediate position in both temperature and ionization timescale. The characteristic timescale inferred for this component (4.4--5.2\,kyr) is comparable to the Sedov age of the remnant, supporting the interpretation that it predominantly represents the diffuse ICM heated by the  
blast shock and constitutes the dominant volume-filling plasma responsible for the bulk of the X-ray emission.

In addition to the differences in physical origin, the relationship between $n_e t$ and $kT_e$ may also be affected by electron-proton (e-p) temperature non-equilibrium behind collisionless shocks. Immediately after shock heating, the energy is not equally partitioned between electrons and protons, and the electron temperature can be substantially lower than the proton temperature, particularly for high-Mach-number shocks. Subsequent Coulomb collisions and plasma processes gradually transfer energy from protons to electrons, driving the system toward thermal equilibrium. If the plasma has not yet reached full e-p equilibration, the inferred electron temperature may not directly trace the total post-shock energy content, potentially introducing additional scatter or apparent trends in the $n_et-kT_e$ relation.

In this context, the single power-law relation between $n_e t$ and $kT_e$ adopted in the DEM modeling should be regarded as a simplified empirical parameterization rather than a physically unique description \citep{Sun2025}. In a multi-phase medium with different shock histories, densities, and equilibration timescales, deviations from a single power-law behavior are expected. Therefore, while the DEM results capture the overall temperature distribution of the X-ray-emitting plasma, the detailed $n_et-kT_e$ correlation should be interpreted with caution.

We note that the NEI model fitting exhibits intrinsic degeneracies, particularly between the electron temperature ($kT_e$) and ionization timescale ($n_et$), as also indicated by our MCMC analysis. Such correlations are partly model-dependent, reflecting the flexibility of NEI models in reproducing similar spectral features with different parameter combinations. Therefore, caution is required when interpreting these parameters physically, and the derived plasma conditions should be considered tentative within the model framework.

\subsection{Progenitor}\label{sec:progenitor}
To investigate the nature of the progenitor star, we first derived a representative set of elemental abundances for the entire remnant.
Given the significant spatial variations, a simple average of the abundances from each region is not appropriate. Instead, we took the mass-weighted elemental abundance for each element X, defined as ${\rm{X}} = \Sigma m_i{\rm X}_i/\Sigma m_i$, where $m_i$ and ${\rm X}_i$ are the gas mass and the abundance of element X in the $i^{th}$ region. Based on the best-fit values from the \textit{Chandra} WVT analysis, we calculated these mass-weighted abundances 
and then determined their standard deviations, which we adopted as the errors. 
For the observational results based on the XMM-\textit{Newton} data, we directly applied error propagation formulas to compute the abundance ratios.

The abundance fitting results indicate that many regions show no significant overabundances. This suggests that the high-temperature component also contains a substantial amount of ISM and cannot be entirely attributed to pure ejecta. 
To account for this dilution effect in the nucleosynthesis modeling, we adopted the method of \citet{Weng2022}, in which the ambient ISM is mixed with the metal-rich ejecta when calculating the expected abundance. In our case, the total mass of the mixed ejecta and ISM is constrained by our previous spectral analysis, $\sim330\,M_{\odot}$. 
As explained in \citet{Weng2022}, mixing a large amount of ISM with initially overabundant ejecta naturally drives the elemental abundance ratios toward 1, while preserving the relative abundance trends.

We compared the observed abundance ratios relative to Si with the predictions from SN nucleosynthesis models. The abundance of element X relative to Si is defined as $\rm Z_X/Z_{Si}$, where $\rm Z_X$ is the abundance ratio of element X relative to its solar value.
We adopted the yields for massive stars from \citet{Sukhbold2016}, 
covering a zero-age main-sequence mass range of 9.0--120\,$\rm M_{ \odot}$. 
Specifically, we utilized the yields from the N20 and W18 central engine models,
which include contributions from both explosive ejecta and stellar winds.

\begin{figure}
    \centering
    \subfigure[]{\includegraphics[width=0.4\textwidth]{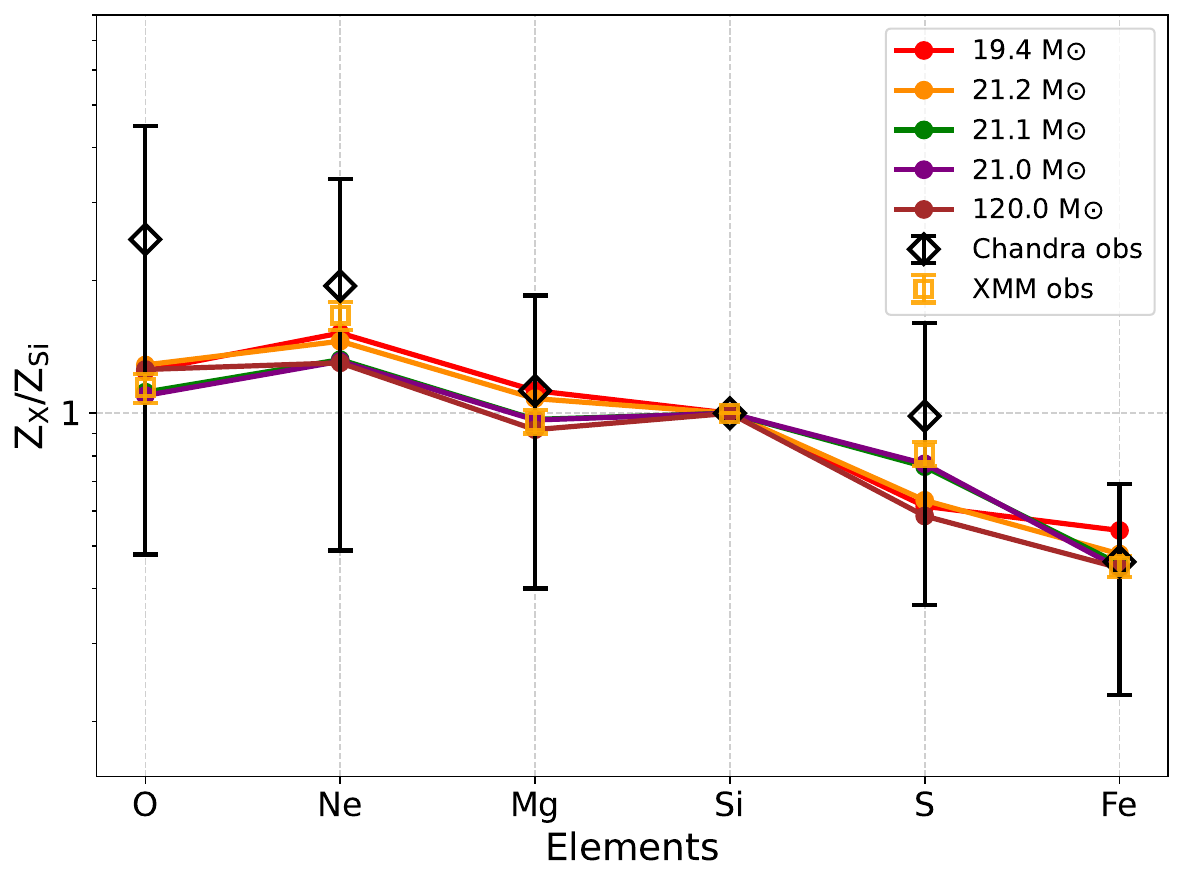}}
    \subfigure[]{\includegraphics[width=0.4\textwidth]{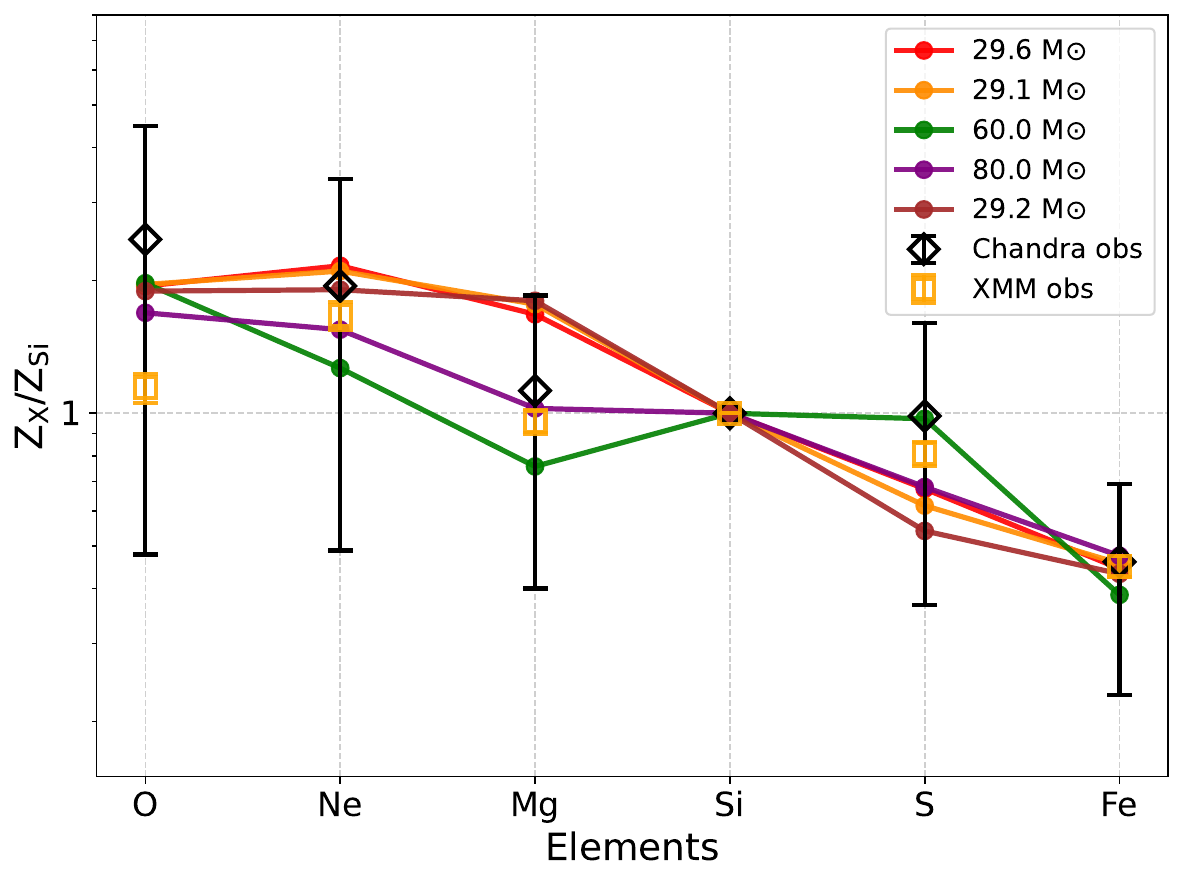}}
    \subfigure[]{\includegraphics[width=0.4\textwidth]{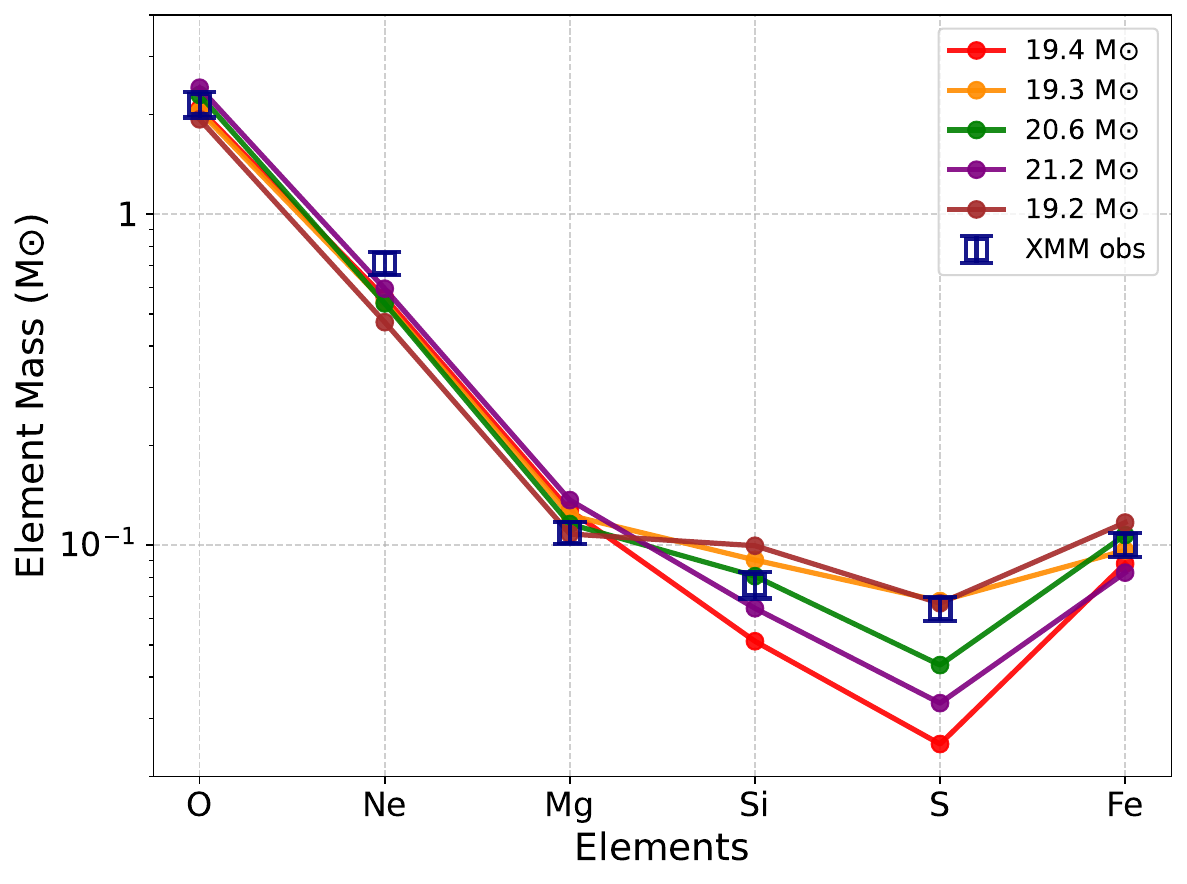}}
    
    \caption{Comparisons between element abundance ratios and SN models (N20, \citealt{Sukhbold2016}). 
    Black hollow diamonds and error bars represent the results from the \textit{Chandra} observation, orange hollow square boxes represent the results from the XMM-\textit{Newton} observation, and the colored lines represent the product abundance ratios for progenitors of  
    various masses.  
    (a) Five sets of models that best match the results from XMM-\textit{Newton} 
    observation. (b) Five sets of models that best match the 
    results from the \textit{Chandra} observation. (c) Comparison of individual elemental masses with the N20 model.}
    \label{fig:N20}
\end{figure}

\begin{figure}
    \centering
   
    \subfigure[]{\includegraphics[width=0.4\textwidth]{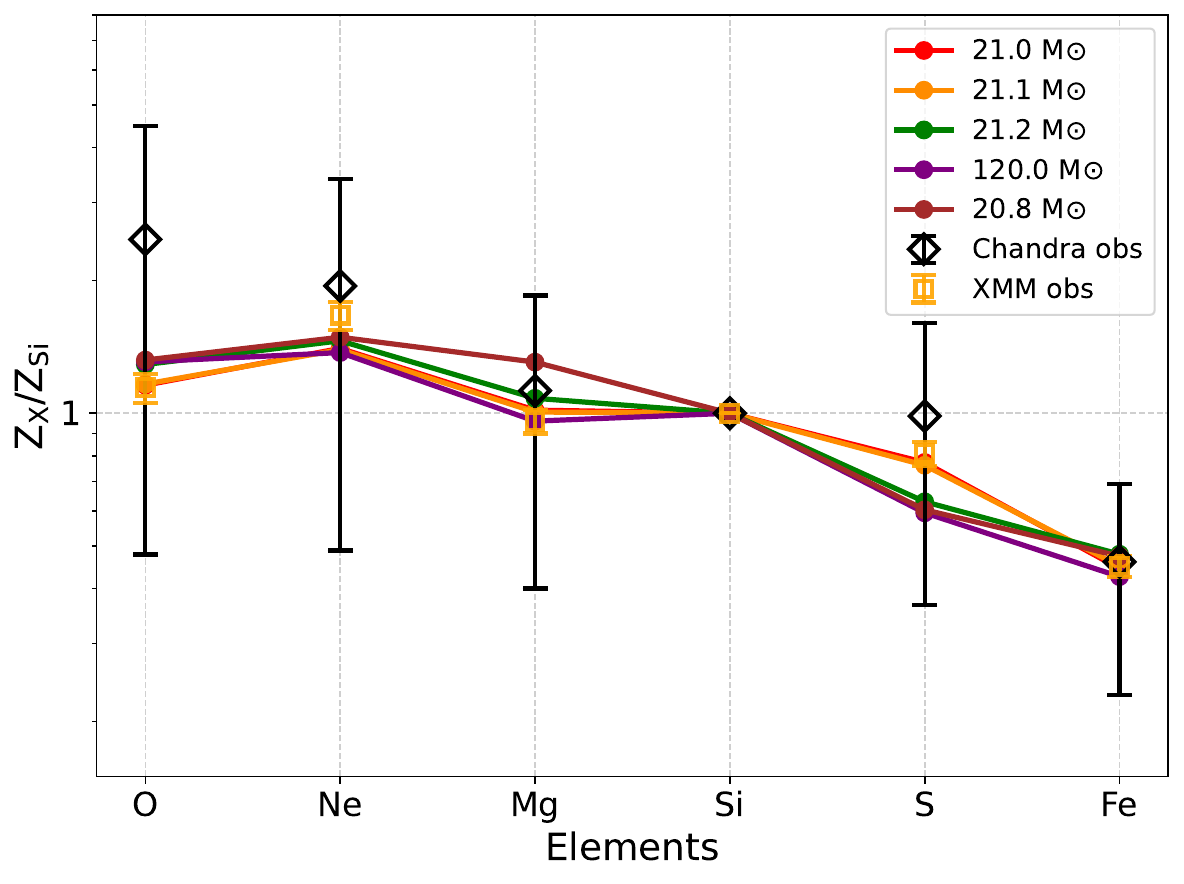}}
    \subfigure[]{\includegraphics[width=0.4\textwidth]{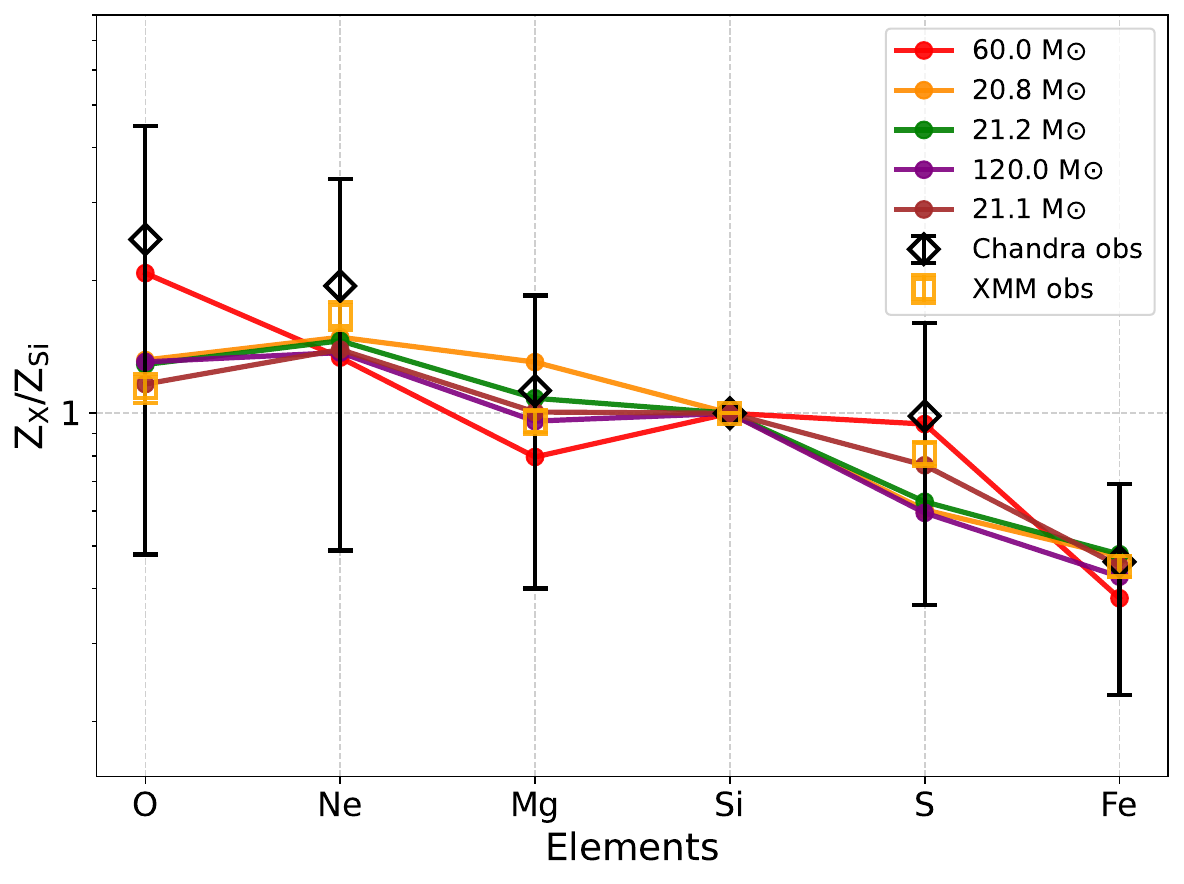}}
    \subfigure[]{\includegraphics[width=0.4\textwidth]{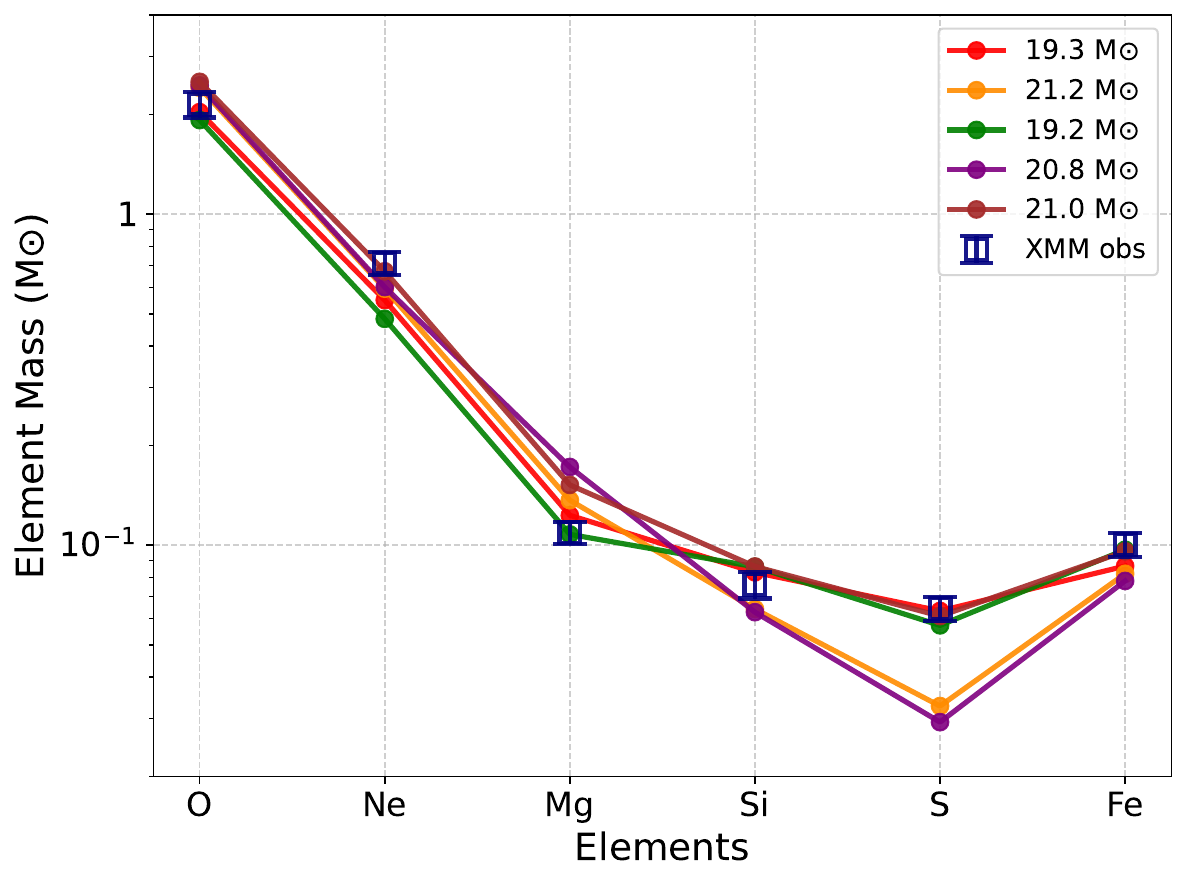}}

    \caption{Comparisons between abundance ratios and SN models (W18, \citealt{Sukhbold2016}). Black hollow diamonds and error bars represent the results from the \textit{Chandra} observation, orange hollow square boxes represent the results from the XMM-\textit{Newton} observation, and the colored lines represent the product abundance ratios for progenitors of various masses.  
    (a) Five sets of models that best match the XMM-\textit{Newton} observation. (b) Five sets of models that best match the results from the \textit{Chandra} observation. (c) Comparison of individual elemental masses with the W18 model.}
    \label{fig:W18}
\end{figure}

As shown in Figure \ref{fig:N20} and \ref{fig:W18}, 
the O/Si, Ne/Si, and Mg/Si ratios derived from the \textit{Chandra} WVT analysis have relatively large uncertainties, whereas those from the XMM-\textit{Newton} observation are more tightly constrained. 
Overall, the abundance ratios from the two datasets are consistent with each other, with the exception of the O/Si ratio, where the most significant discrepancy is observed. However, the XMM-\textit{Newton} data points still fall within the error bars of the \textit{Chandra} results. 

We 
then considered and compared both datasets against the models.
The curves in Figure \ref{fig:N20}a and Figure \ref{fig:W18}a represent all model predictions from the N20 and W18 engines in \citet{Sukhbold2016} that fall within the 
uncertainty range of the results from the \textit{Chandra} observation. Most of these correspond to progenitor stars with masses below 30\,$M_{\odot}$. When considering the results from the XMM-\textit{Newton} and \textit{Chandra} 
observations separately, the progenitor mass most consistent with the XMM-\textit{Newton} observation lies in the range of 19--22\,$M_{\odot}$. In contrast, the \textit{Chandra} observation allows for additional solutions, including very massive progenitors exceeding 60\,$M_{\odot}$, alongside the $\sim$ 20--30\,$M_{\odot}$ solution. 
Nevertheless, both datasets are consistent with a progenitor mass in the range of $\sim$19--22$\,M_{\odot}$. We note, however, that the \textit{Chandra} data also permit a higher-mass solution ($\gtrsim40 M_{\odot}$), and the available constraints do not allow us to definitively exclude either possibility.

\begin{figure}[h]
    \centering
    \subfigure[]{\includegraphics[width=0.4\textwidth]{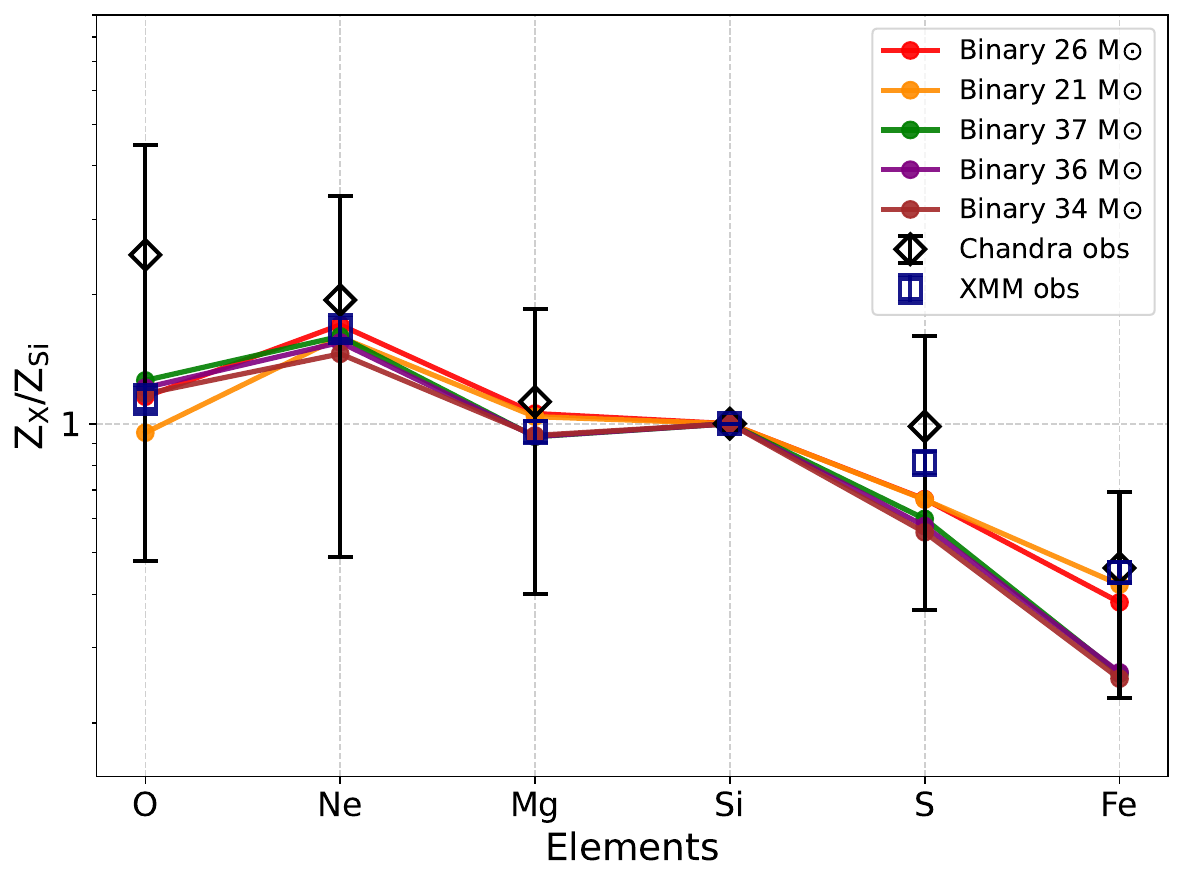}}
    \subfigure[]{\includegraphics[width=0.4\textwidth]{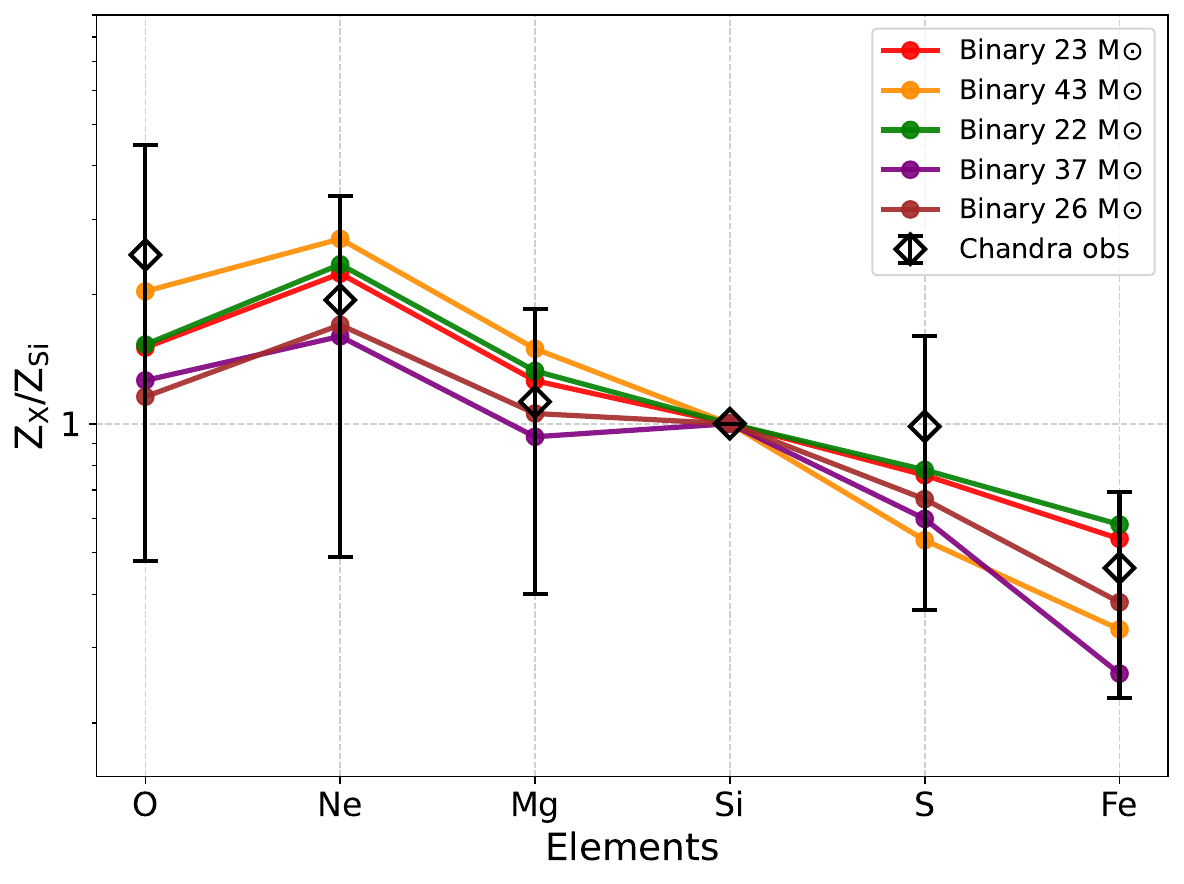}}
    \subfigure[]{\includegraphics[width=0.4\textwidth]{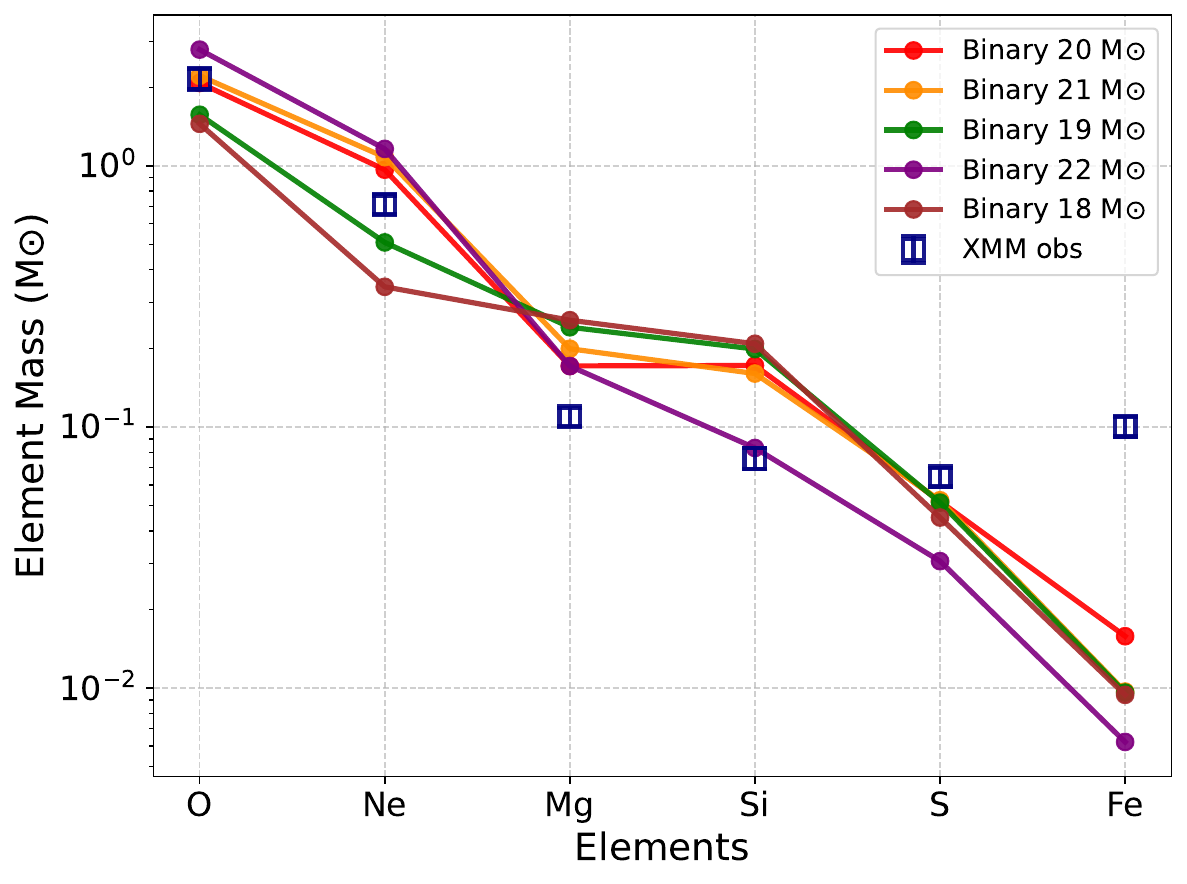}}

    \caption{Comparisons between abundance ratios and stripped-binary SN models \citep{Farmer2023}. Black hollow diamonds and error bars represent the results from \textit{Chandra}, darkblue hollow square boxes represent the results from XMM-\textit{Newton}, and the colored lines represent the product abundance ratios for progenitors of different masses. (a)  Five sets of models that best match the XMM-\textit{Newton} results. (b) Five sets of models that best match the \textit{Chandra} results. (c) Comparison of individual elemental masses with the stripped-binary SN model. }
    \label{fig:binary}
\end{figure}

We also considered the scenario of progenitors in binary systems. 
Using the nucleosynthetic yields for stripped-envelope stars in binary systems from \citet{Farmer2023}, we performed 
similar comparison. As shown in Figure \ref{fig:binary}, a few of the binary models satisfactorily reproduce the observed abundance ratios of N63A, with the corresponding binary mass distribution being relatively scattered, but the most favorable solutions clustering around $20\,M_{\odot}$.

To further discriminate between the progenitor scenarios, we performed an alternative analysis using the global fit results of XMM-\textit{Newton} observation. For each element (O, Ne, Mg, Si, S, Fe) that shows potential enhancement, we calculated its 
mass in the ejecta. The mass of an element in the ejecta, $M_X$, was estimated by isolating the 
contribution above the LMC ISM level, using the formula $M_X = (Z_X^{fit}-Z_X^{LMC})\times (N_X^{\odot}/N_H^{\odot})\times M_H/m_H \times A_X$, where $Z_X^{fit}$ is the abundance from spectral fitting, $Z_X^{LMC}$ is the standard LMC ISM abundance, $(N_X^{\odot}/N_H^{\odot})$ is the solar abundance ratio, $M_H(\approx 330\,M_{\odot})$ is the total hydrogen mass of 
component 2 in our 
spectral fit, $m_H$ is the mass of a hydrogen atom, and $A_X$ is the atomic mass of the element X in units of the hydrogen atom mass. 
This calculation quantifies the mass of each element within the ejecta of component 2 that is responsible for the observed abundance enhancements. 
We then compared these derived ejecta masses with the predictions from single-star \citep{Sukhbold2016} and binary-stripped \citep{Farmer2023} nucleosynthesis models. The comparison, shown 
panel d in Figure \ref{fig:binary}), reveals that the ejecta masses derived from XMM-\text{Newton} are broadly consistent with the yields from single-star models with a progenitor mass of approximately $20\,M_{\odot}$. In contrast, the binary-stripped models generally predict ejecta masses that do not simultaneously match the observed masses of O, Ne, Mg, and Si in the ejecta of N63A.

Previous studies have suggested a progenitor mass of approximately 40\,$\rm M_{\odot}$ \citep{Karagoz2023}, which would imply a remarkably massive star. This scenario remains plausible when considering the presence of massive stars in the OB association linked to N63A. However, stars of such  
a large mass face significant challenges. On one hand, they experience severe mass loss via stellar winds, likely shedding most of their mass and ending their evolution with only a few solar masses before exploding as Type Ib or Ic SNe \citep{Sukhbold2016}. 
On the other hand, alternatively, they may avoid SN explosions entirely and directly collapse into black holes \citep{Heger2003}. This scenario is still poorly represented in most existing models.  
Considering that N63A is located in the LMC, where the ISM has  
relatively low metallicity, the extent to which this environment influences nucleosynthetic yields remains uncertain. 
Consequently, we cannot rule out a progenitor mass $\gtrsim40 M_{\odot}$ based on present data. Our analysis indicates that a mass of $\sim 20\,\rm M_{\odot}$ provides a satisfactory match to the observed abundance ratios and ejecta masses, but we emphasize that a higher-mass origin remains an open possibility.

\section{Summary}\label{conclusion}
We performed a comprehensive X-ray spectroscopic study of the SNR N63A in the LMC, utilizing both \textit{Chandra} and XMM-\textit{Newton} observations. Our analysis yields new insights into its plasma properties, chemical composition, progenitor system, and evolutionary history of the remnant.
The main results are summarized as follows:
\begin{itemize}
\item[1.] 
The global X-ray spectrum of N63A is best described by a model comprising three distinct thermal plasma components with temperatures  
around 0.3, 0.7, 1.5\,keV, respectively. This temperature structure arises from the SNR's interaction with a multi-phase ISM. The low-temperature component originates from the evaporation of engulfed dense clumps. The intermediate-temperature component is dominated by shock-heated ICM mixed with ejecta, and the high-temperature component is likely produced by  
the shocks reflected by the clumps. 
\item[2.] A DEM analysis reveals a peak temperature distribution around 0.7\,keV and a  
positive dependence of ionization timescale on temperature. This suggests a complex relationship between plasma temperature and ionization history, potentially influenced by electron-ion temperature equilibrium and other cooling processes.
\item[3.] Spatially-resolved spectral analysis using \textit{Chandra} data reveals a complex distribution of elemental abundances. Enhanced abundances of O, Ne, and Mg in the central regions likely trace the distribution of ejecta. The ionization timescale is significantly higher in the eastern (shock-ionized) optical lobes than in the western (photoionized) lobe, consistent with the optical classifications. 
\item[4.] By comparing the observed abundance ratios with 
the predictions from core-collapse nucleosynthesis models, we find that the progenitor mass is most consistent with $\sim20\,M_{\odot}$.
This favors a single-star origin over a 
scenario involving a stripped progenitor in a binary system.

\end{itemize}

\begin{acknowledgments}
This research employs a list of \textit{Chandra} datasets, obtained by the \textit{Chandra} X-ray Observatory, contained in~\dataset[DOI: 10.25574/00777]{https://doi.org/10.25574/00777}.
The authors wish to thank Gao-Yuan Zhang and Yi-Heng Chi for helpful discussions. H.C. acknowledges Li Ji for supervising the undergraduate course project which motivated this study, and Zixin Wei, Tian-Xian Luo, and Wen-Yu Fan for their collaboration during that early-stage course project, which helped shape the initial direction of this work.
This work is supported by the National Natural Science Foundation of China (NSFC) under grants 12121003, 12573047, 12393852,  and 12503030. 

\textit{Software:} ATOMDB \citep{Smith2001,Foster2012}, CIAO \citep[vers. 4.16,][]{Fruscione2006}, WVT binning \citep{Cappellari2003,Dickey1990}, SAS\citep[vers. 20.0, ][]{Gabriel2004}, XSPEC \citep{Arnaud1996}, SPEX \citep{Kaastra1996}, DS9\footnote{\url{http://ds9.si.edu/site/Home.html}} \citep{Joye2003}.
\end{acknowledgments}

\bibliography{ref}{}
\bibliographystyle{aasjournalv7}

\appendix
\section{Examples of Fitted \textit{Chandra} Spectra}
Three examples of \textit{Chandra} spectra and their fitting residuals of \texttt{tbabs*tbvarabs*(vnei+vnei)} model are shown below.
\begin{figure*}[h]
    \centering
    \subfigure[]{\includegraphics[width=0.45\textwidth]{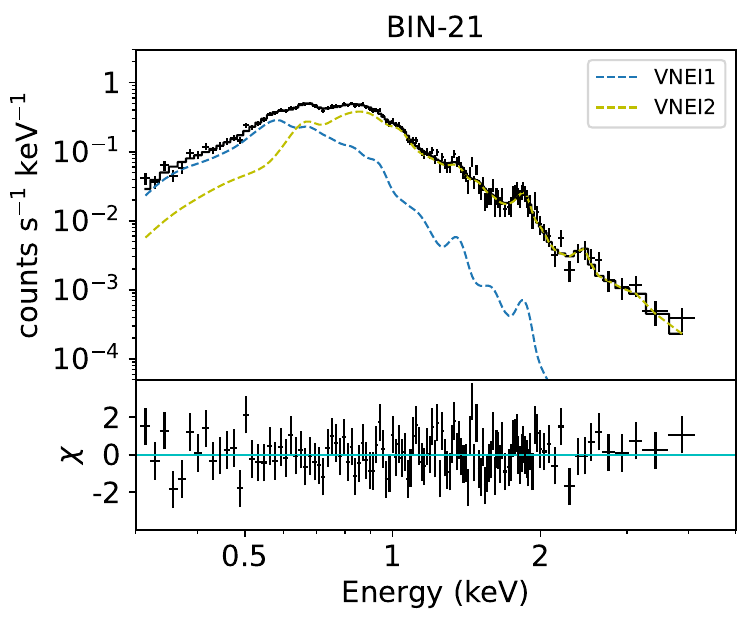}}
    \subfigure[]{\includegraphics[width=0.45\textwidth]{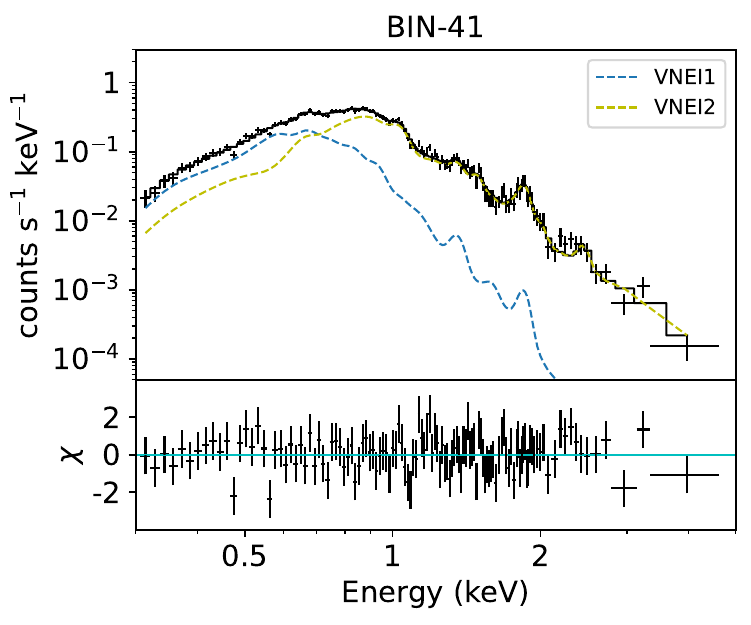}}
    \subfigure[]{\includegraphics[width=0.45\textwidth]{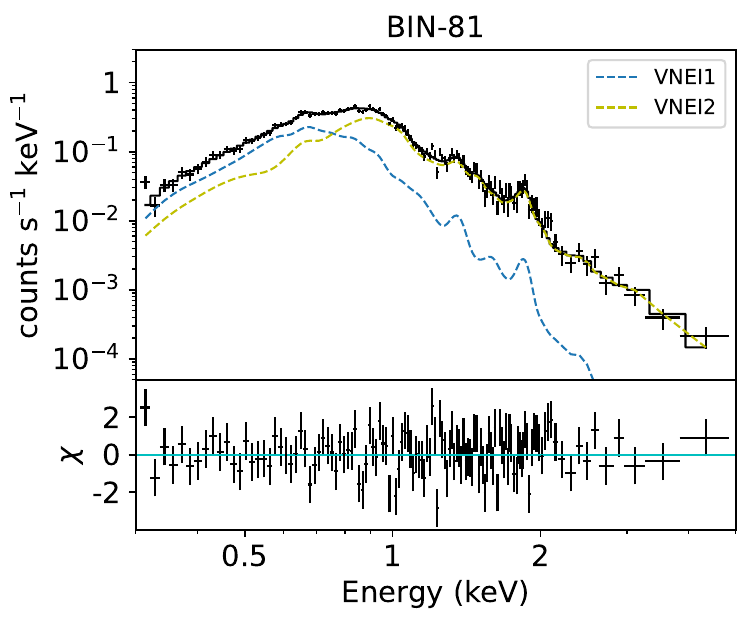}}
    \subfigure[]{\includegraphics[width=0.45\textwidth]{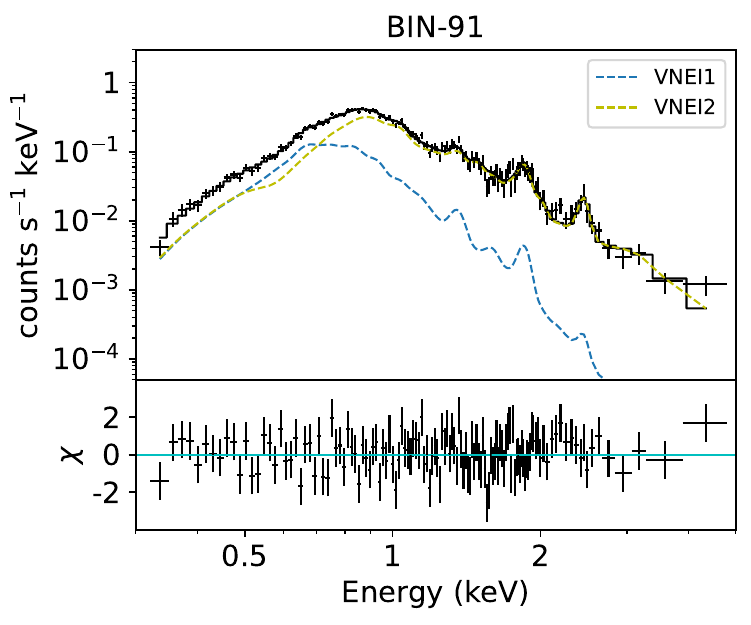}}
    
    \caption{Examples of fitted spectra from three tessellated bin regions. The data points are the background-subtracted spectra, and the thin solid lines are their best-fit models. Different model components are plotted as dotted lines with different colors.}
    \label{fig:wvt_spec}
\end{figure*}

\section{Fitted XMM-\textit{Newton} EPIC Spectra with Background}
The XMM-\textit{Newton} background spectra of N63A were extracted from an annular region surrounding the source (Figure \ref{fig:xmm_image}. The background was modeled using three components: the Local Hot Bubble (LHB), Galactic halo (GH) emission, and the unresolved cosmic X-ray background (CXB) dominated by active galactic nuclei (AGNs). The temperature of the LHB component was fixed at 0.1\,keV, and the normalization was fixed at $10^{-5}\,\rm cm^{-5}$. The GH and other foreground emissions were modeled using two absorbed \texttt{apec} components with temperatures of approximately 0.26\,keV and 0.89\,keV, respectively. The AGN background was represented by a power-law model with a photon index of $\Gamma=1.4$, together with an additional LMC absorption component assuming an LMC metallicity of 0.5 solar. For the instrumental background, the MOS spectra have two gaussian lines, the Al-K (1.49\,keV) and Si-K (1.75\,keV) fluorescence lines, while pn also mainly has two lines, the Al K$\alpha$ (1.49\,keV) and a complex of Ni-K$\alpha$, Cu-K$\alpha$, Zn-K$\alpha$ lines around 8\,keV. An additional Gaussian component near 1.3\,keV was included to account for residual features. 
\leavevmode
\begin{figure*}[h!]
    \centering
    \includegraphics[width=0.5\textwidth]{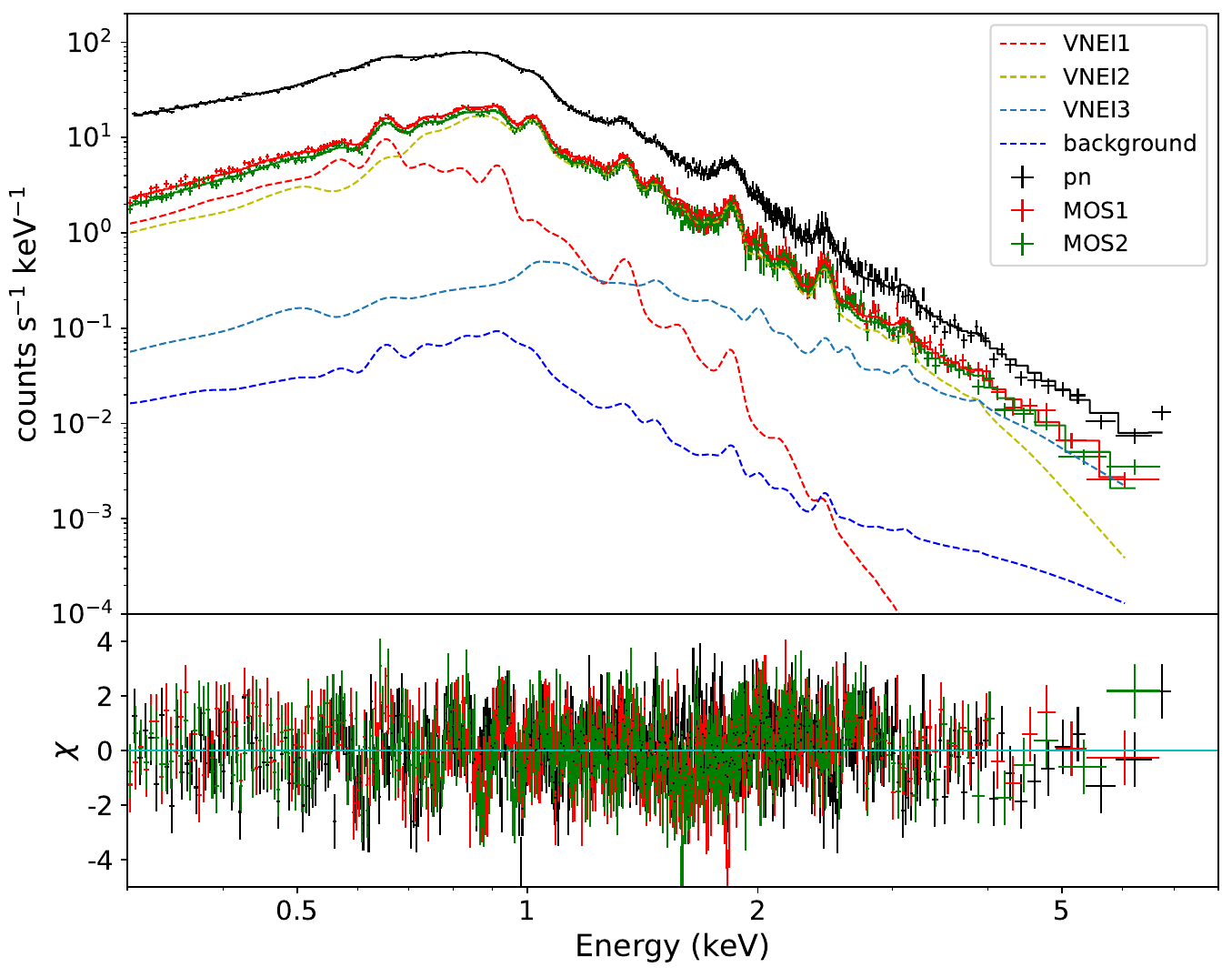}
    \caption{XMM-\textit{Newton} EPIC spectrum of N63A, extracted without background subtraction. The data were fitted with a model that incorporates the background component. The individual model components (including the source emission and the background) are indicated by the dashed lines in different colors.}
    \label{fig:spec_3vnei_bkg}
\end{figure*}

\section{\texorpdfstring{Results of MCMC}{Results of MCMC}}
\begin{figure*}[h!]
    \centering
    \subfigure[]{\includegraphics[width=0.45\textwidth]{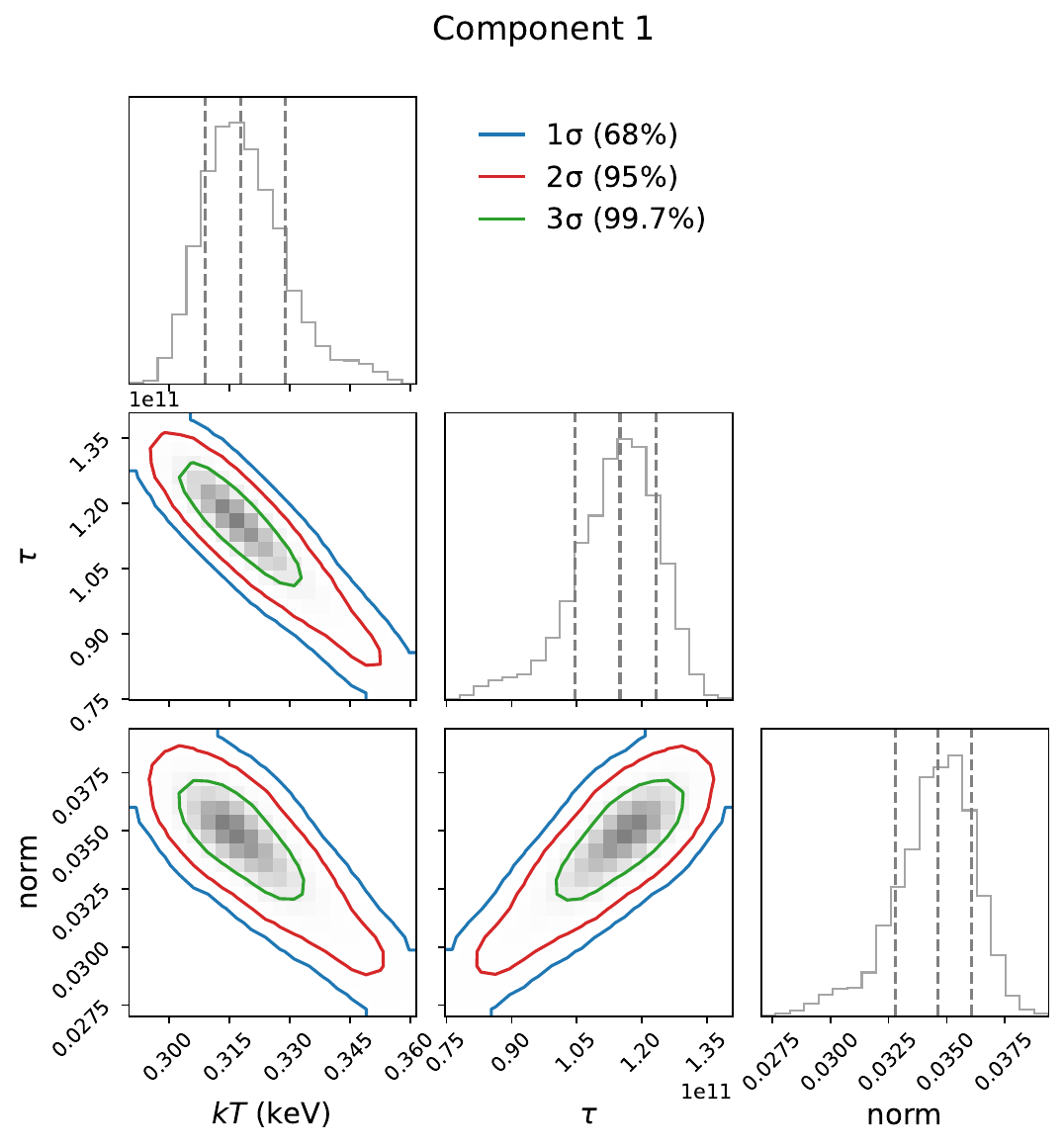}}
    \subfigure[]{\includegraphics[width=0.45\textwidth]{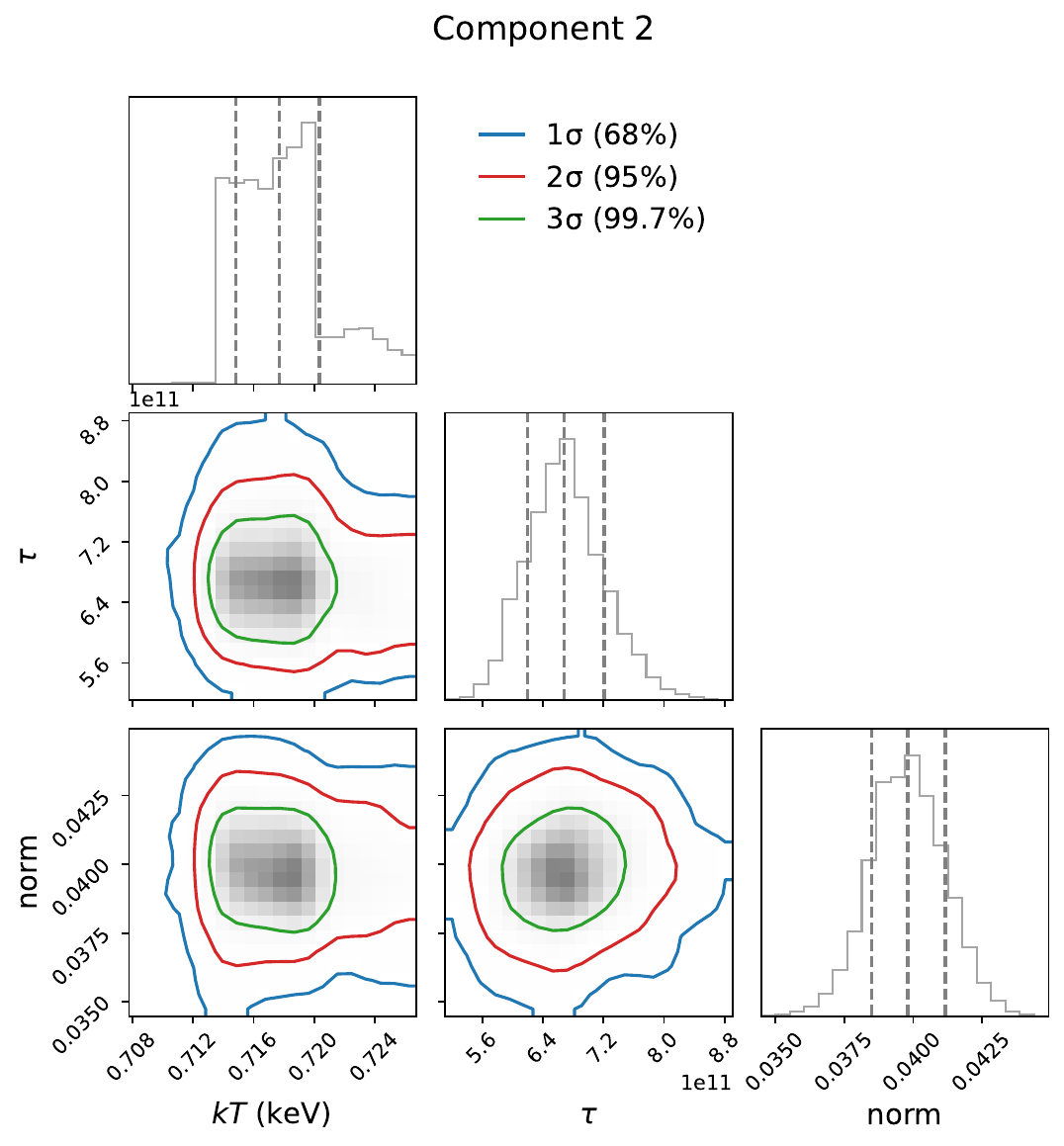}}
    \subfigure[]{\includegraphics[width=0.45\textwidth]{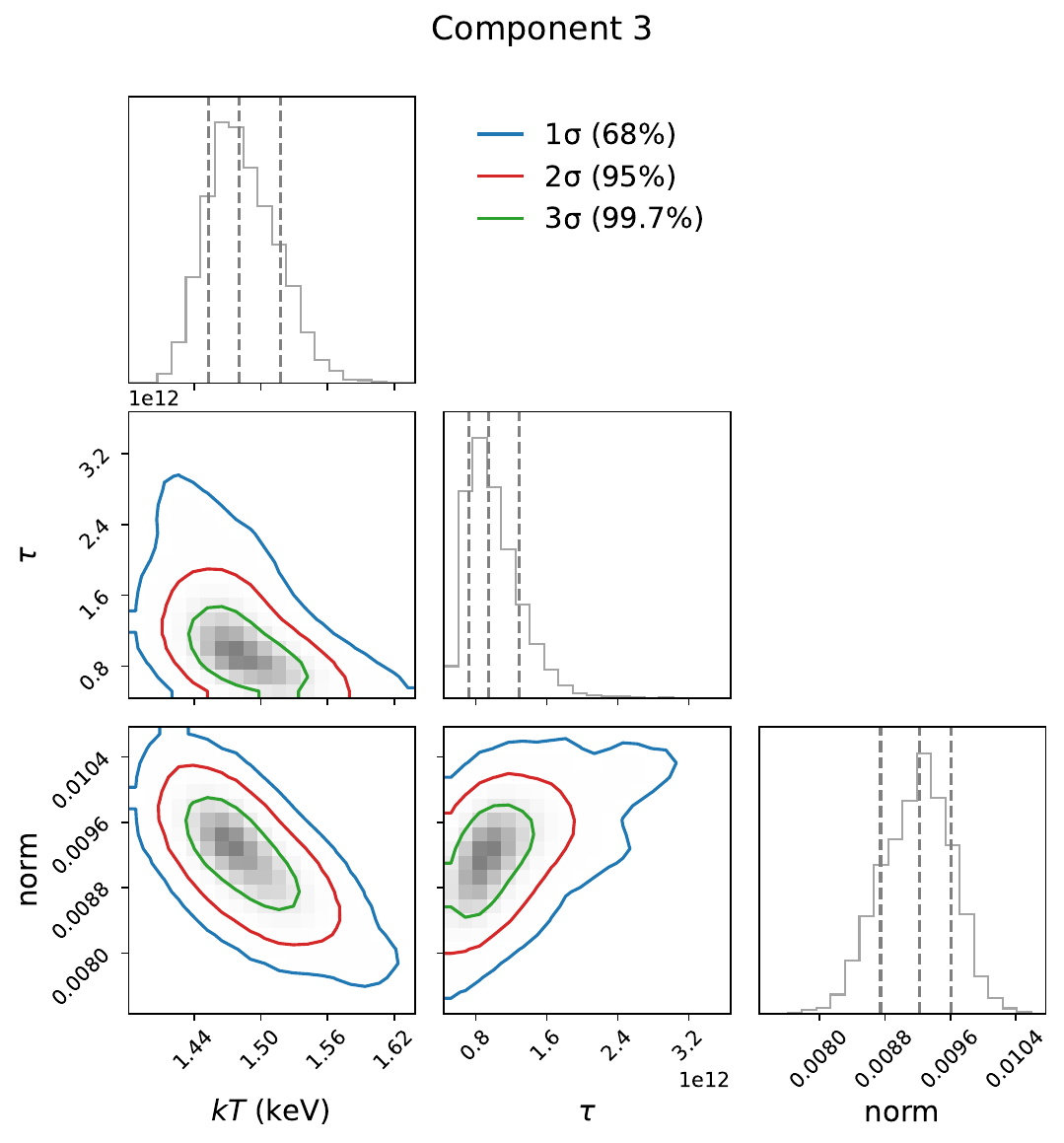}}
    
    \caption{Corner plots of the MCMC posterior distributions for the three plasma components. Panels (a), (b), and (c) correspond to the low- (Component 1), intermediate- (Component 2), and  high-temperature (Component 3) components, respectively. In each panel, the two-dimensional contours show the 68\%($1\sigma$), 95\%($2\sigma$), 99.7\%($3\sigma$) confidence regions for pairs of parameters. The one-dimensional histograms display the marginal distributions with dashed lines marking the 16th, 50th, and 84th percentiles. The contour colors are: blue (1$\sigma$), red (2$\sigma$), and green (3$\sigma$).}
    \label{fig:mcmc_corner}
\end{figure*}

The corner plots (Figure \ref{fig:mcmc_corner}) show the joint posterior distributions of temperature, ionization timescale, and normalization for the three components obtained from the MCMC analysis. For the low-temperature component (Component 1), the two-dimensional contours appear elongated, indicating significant degeneracies among temperature, ionization timescale, and normalization. The distribution of the intermediate-temperature component is nearly circular with no obvious correlation, suggesting that $kT_2$ and $\tau_2$ are well constrained independently. The high-temperature component exhibits a moderately elliptical contour with a moderate correlation, implying partial degeneracy among its parameters. This difference likely reflects the varying contributions and diagnostic information content of the different plasma component in the spectra. Component 2 may be the brightest or possess the richest diagnostic features, which is also consistent with DEM results. Component 1 probably originates from a more diffuse or mix region, making it difficult to separate temperature and ionization state.

\end{document}